\documentclass[prb,aps,amsmath,amssymb,twocolumn,superscriptaddress]{revtex4-2}

\usepackage{graphicx}  % Include figure
\usepackage{textcomp}
\usepackage{gensymb}
\usepackage{color}
\usepackage{placeins}  % Add this in your preamble
\usepackage{xcolor}
\usepackage{float}

\graphicspath{{figs}}
\usepackage[colorlinks=true, citecolor=blue, linkcolor=blue, urlcolor=blue]{hyperref}
\hypersetup{colorlinks=true, citecolor=blue, linkcolor=blue, urlcolor=blue}
\makeatletter
\AtBeginDocument{%
        \renewcommand{\@biblabel}[1]{[#1]} % Maintain APS citation format
}
\makeatother

\definecolor{BluBondi}{rgb}{0.00,0.58,0.71}
\definecolor{tangerine}{rgb}{0.944,0.522,0}
\definecolor{brown}{rgb}{0.633,0.156,0.156}
\definecolor{teal}{rgb}{0,0.502,0.502}

\newcommand{\editor}[2]{%
  \expandafter\newcommand\csname #1note\endcsname[1]{%
    \textcolor{#2}{[\textbf{#1:}  ##1]}}%
  \expandafter\newcommand\csname #1\endcsname[1]{%
    \textcolor{#2}{##1}}%
  \expandafter\newcommand\csname #1cancel\endcsname[1]{%
    \textcolor{#2}{\sout{##1}}}%
  \expandafter\newcommand\csname #1change\endcsname[2]{%
    \textcolor{#2}{\sout{##1} ##2}}%
  \newenvironment{#1text}{\color{#2}}{\color{black}}
}
\editor{KD}{BluBondi}
\editor{TS}{teal}
\editor{SN}{brown}
\editor{FM}{tangerine}

\newcommand{\lcc}{Laboratoire Charles Coulomb (L2C), UMR 5221 CNRS-Université de Montpellier, Montpellier, France.}

\begin{document}

\title{Electronic transport in BN-encasulated graphene limited by remote phonon scattering}

\author{K. Dinar}
    \affiliation{\lcc}
    \affiliation{Aix Marseille Univ, CNRS, CINaM, AMUTech, Marseille, France.}

\author{F. Macheda}
\affiliation{Dipartimento di Scienze e Metodi dell’Ingegneria,
University of Modena and Reggio Emilia, Reggio Emilia, Italy}
\affiliation{Dipartimento di Fisica, Sapienza Università di Roma, Roma, Italy}
\author{A. Guandalini}
\affiliation{Dipartimento di Fisica, Sapienza Università di Roma, Roma, Italy}
\author{M. Paillet}
\affiliation{\lcc}
\author{C. Consejo}
\affiliation{\lcc}
%\author{G. Sigu}
%\affiliation{\lcc}
\author{F. Teppe}
\affiliation{\lcc}
\author{B. Jouault}
\affiliation{\lcc}
\author{T. Sohier}
\affiliation{\lcc}
\author{S. Nanot}
\affiliation{\lcc}

\begin{abstract}
We study the impact of BN's phonons on the electrical resistivity of hBN-encapsulated graphene. While encapsulation yields high-mobility devices, the surrounding BN itself introduces remote scattering from polar optical phonons, whose role in standard resistivity measurements remains unclear. We combine high-quality transport experiments with ab initio calculations including a proper treatment of dynamically screened remote interactions. We demonstrate that hBN's out-of-plane phonons strongly influence resistivity between 150 K and room temperature, whereas higher-energy LO modes and intrinsic graphene phonons alone cannot explain the observed trends. The coupling between electrons and the BN's phonons becomes more pronounced at low carrier densities due to reduced screening. Our findings establish that remote phonon scattering fundamentally limits transport in encapsulated graphene, solving a longstanding debate.
\end{abstract}

        \maketitle

Encapsulation is the process of placing a 2D material in between nanomembranes of protective materials, typically multilayers of hexagonal Boron Nitride (hBN).
This is generally aimed to preserve the materials' intrinsic properties, protecting it from oxidation and impurities \cite{Dean2010} and suppressing intrinsic roughness \cite{PhysRevB.96.014101}. In the context of electronic transport, the benefit of encapsulation is remarkably demonstrated by the fabrication of high-mobility devices from BN-encapsulated graphene with low charge inhomogeneity \cite{Dean2010,Xue2011,Wang2013a}. Nonetheless, eliminating all extrinsic defects comes at a cost. The encapsulator itself can alter the 2D material's transport properties through long-range, field-mediated couplings between the encapsulator's excitations and the material's electrons. For BN-encapsulated graphene, the detrimental excitations are identified with the BN's polar (remote) phonons, which can increase the value of graphene resistivity.

The temperature-dependent resistivity of graphene is mainly determined by intrinsic acoustic phonons at low temperature and optical phonons at higher temperature. The latter include also the phonons of the environment, such as remote interfacial phonons from polar dielectric substrates, which are fundamental to explain the resistivity measurements for graphene grown on SiO$_2$ \cite{Chen2008,Fratini2008,Efetov2010,PhysRevLett.100.016602} or high-$\kappa$ oxides \cite{Konar2010,You2019}, and in ferroelectric-gated graphene \cite{Chen2022}.
 Compared to such substrates, BN-encapsulated graphene devices usually yield lower resistivity \cite{Dean2010,Wang2013a,Banszerus2019,You2019} and are considered closer to the intrinsic performances of isolated graphene. Accordingly, the contribution of BN's optical phonons to graphene resistivity is considered negligible up to room temperature \cite{You2019}. However, high-bias transport measurements indicate that the coupling of electrons to BN's optical phonons is crucial to determine the current saturation \cite{Yang2018,Baudin2020,Guo2025a}, resulting in an apparent inconsistency.

Theoretically, the resistivity of isolated graphene has been extensively studied \cite{Hwang2008,DasSarma2011,PhysRevLett.101.096802} and computed using state-of-the-art ab initio methods \cite{Kaasbjerg2012,Park2014,Sohier2014}.
At low temperature, in the so-called equipartition regime \cite{Sohier2014}, the temperature-dependent resistivity is dominated by acoustic phonons, and behaves linearly as $\rho \propto  T$, matching experimental values \cite{Sohier2014}.
At higher temperatures, the resistivity is mainly determined by graphene's optical phonons located at zone center ($\Gamma$) and zone border (K) \cite{Park2014,Sohier2014}. In particular the A$'_1$ mode near the K point shows a very strong coupling to electrons, as recently hinted by infrared Raman spectroscopy \cite{Venanzi2023} and many-body calculations \cite{Guandalini2025,PhysRevB.77.041409}. On the contrary, the contributions from remote phonons of the environment have only been studied with empirical semi-analytical models. For graphene supported on bulk dielectric substrates, it is shown that remote interfacial phonons dominate high-temperature resistivity \cite{Fratini2008,Konar2010,Perebeinos2010,Schiefele2012,Ong2012,Lin2013}.
For BN-encapsulated graphene, models neglecting screening from graphene's electrons predict significant scattering from BN's polar optical phonons \cite{Perebeinos2010}, while models including screening hint to negligible contributions \cite{Schiefele2012}.
Addressing this discrepancy requires a rigorous and quantitative evaluation of remote BN phonon scattering and its effect on resistivity, based on ab initio calculations that carefully incorporate the screening provided by graphene.

In this work, we investigate the influence of remote electron-phonon coupling on the carrier mobility of monolayer graphene encapsulated in hexagonal boron nitride (hBN), in the high carrier density regime. We fabricated devices that exhibit low charge inhomogeneity and are virtually defect-free, as confirmed by Raman spectroscopy, with negligible strain inhomogeneities. These attributes lead to low residual resistivity, yielding room-temperature electronic mobilities up to $40{,}000\,\text{cm}^2\text{/Vs}$. A detailed analysis of temperature-dependent transport measurements is presented and compared with \emph{ab initio} calculations that incorporate both intrinsic graphene phonons and remote scattering from hBN phonons. Our theoretical framework builds upon recent methodological developments described in a companion paper \cite{accpaper}, enabling a unified treatment of remote coupling and dynamical screening from electrons. Specifically, we compute the linear responses of isolated graphene and hBN monolayers, and subsequently model the encapsulated system as a van der Waals heterostructure. From the mean-field dielectric function of the heterostructure, we compute the dispersions of the collective electrodynamical modes—namely, the hBN in-plane (LO) and out-of-plane (ZO) polar phonons and the graphene plasmon—and their coupling to charge carriers \cite{Macheda2024b}. Crucially, electronic excitations and polar phonons are treated on equal footing, capturing the dynamical screening effects due to graphene's electrons. The resistivity is finally obtained by solving the coupled dynamical Boltzmann equations for electrons and electrodynamical modes \cite{accpaper}.

\begin{figure}
        \centering
        \includegraphics[width=1\linewidth]{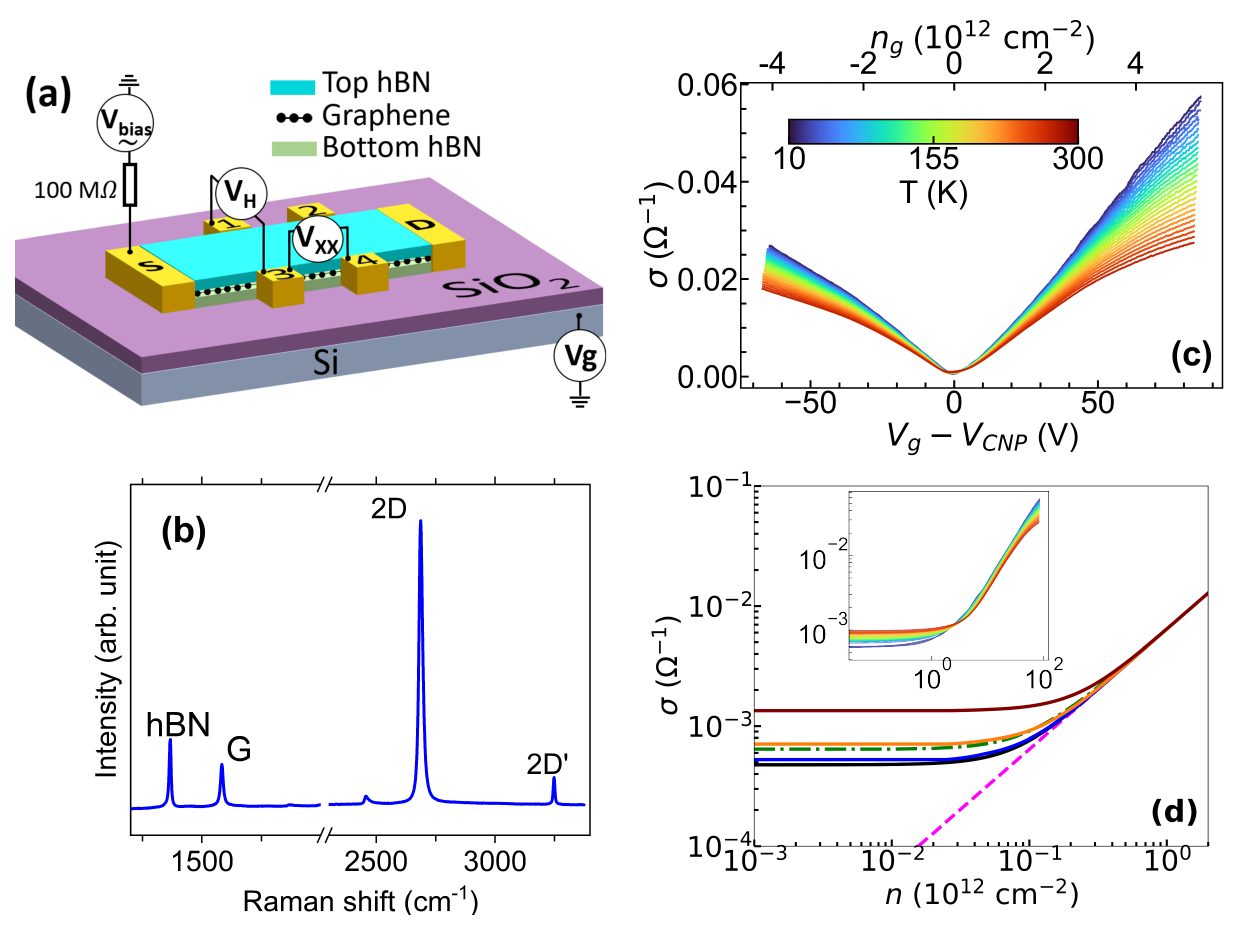}
        	\caption{(a) Schematic illustration of the Hall bar device using an hBN/graphene/hBN stack. (b) Raman spectrum (averaged over the whole area between the contacts 1 to 4) taken with a 532 nm wavelength exitation and showing the characteristic peaks of graphene/h-BN heterostructures. (c) Longitudinal conductivity $\sigma = \sigma_{xx}$ as a function of gate voltage ($V_g$) in the temperature range $10$-$300$ K. $V_{\rm CNP}$ is the reference charge neutrality point voltage, and $n_{g}$ is the carrier density. (d) Different models of the gate voltage  dependence of conductivity for a representative carrier mobility of $40{,}000\,\text{cm}^2\text{/Vs}$. Magenta dashed line: one carrier model. Green dash-dotted line: phenomenological model accounting for a residual charge density $n_0 = 10^{11}\,\text{cm}^{-2}$. Solid lines: thermally activated carriers at $0$, $50$, $150$, and $300$ K, (black, blue, orange, and brown respectively) for a disorder with potential fluctuations\cite{PhysRevB.84.115442} of $35 \,\text{meV}$ (which gives $n_0(0\,\text{K}) \simeq 0.74\,\times 10^{11}\,\text{cm}^{-2}$ and $n_0(300\,\text{K}) \simeq 2.08\,\times 10^{11}\,\text{cm}^{-2}$). The inset of (d) shows the same experimental data as (c) for electron doping ($n_g>0$) on a log-log scale.}
        \label{fig:figure2a}
\end{figure}

\textit{Device fabrication and measurements---}
Hall bar devices of monolayer graphene encapsulated in hexagonal boron nitride (h-BN) have been fabricated using dry-transfer and e-beam lithography on top of Si/SiO$_2$ substrates.
These samples have been measured in closed-cycle helium cryostat equipped with a $1\,\text{T}$ electro-magnet using lock-in technique in order to extract the longitudinal and transverse resistivity in the 10 to 300 K range as a function of backgate voltage applied through the Si substrate.
A schematic representation of the device configuration and measurement are shown in Fig. \ref{fig:figure2a}\textcolor{blue}{(a)} and details are given in the End Matter \ref{app:endmatter}.

In addition, after the electrical measurements, we performed a Raman mapping of the device at 532 nm wavelength. The Raman spectra display the characteristic G peak and a sharp 2D peak. The averaged spectrum over the entire area of interest is shown in figure \ref{fig:figure2a}\textcolor{blue}{(a)}.
The G and 2D bands are located at $\omega_{\rm G} = 1583.7$ cm$^{-1}$ and $\omega_{\rm 2D} =2687.3$ cm$^{-1}$ in average with standard deviation (S.D.) of $0.5$ and $0.9$ cm$^{-1}$, respectively. Their full width at half maximum are $\Gamma_{\rm G} = 13.6$ cm$^{-1}$ and $\Gamma_{\rm 2D} = 17.4$ cm$^{-1}$ in average with S.D. of $1.2$ and $0.8$ cm$^{-1}$, respectively. These results are indicative of a homogeneous monolayer graphene, with weak residual doping and compressive strain as well as small nanometre-scale strain variations from the encapsulation by h-BN~\cite{Neumann2015,PhysRevX.15.021043}.

As shown in Fig. \ref{fig:figure2a}\textcolor{blue}{(c)}, the longitudinal conductivity, $\sigma$, was measured every 10 K as a function of the backgate voltage $V_{\rm{g}}$. It is shown here as a function of $\Delta V_{\rm{g}} = (V_{\rm{g}}-V_{\rm{CNP}})$, where $V_{\rm{CNP}}$ corresponds to the gate voltage at minimum conductivity for each temperature.
and the charge neutrality point of graphene (consistent with the voltage obtained from $n_{\rm{H}} = 0\,\text{cm}^{-2}$).
In order to properly assess the doping dependence and that transport is dominated by carrier scattering, one has to be confident in the extracted carrier density and inhomogeneity. The carrier density, $n_{\rm{g}}$, is calculated using a gate capacitance of $C_{\rm{g}} = 10^{-4}\,\text{F.m}^{-2} $, obtained with a plane capacitor model including a thickness of 300 nm of SiO$_2$ and $30\,\text{nm}$ of bottom hBN (top axis of Fig. \ref{fig:figure2a}\textcolor{blue}{(c)}).
Classical Hall effect was also used to determine the carrier density $n_{\rm H}$ every 50 K and was consistent with the previous value regarding the slope of $n_{\rm H}(V_{\rm{g}})$ and the position of $V_{\rm{CNP}}$, confirming that the minimum conductivity occurs at the charge neutrality point where $n_{\rm{H}} = 0\,\text{cm}^{-2}$.
At room temperature ($T = 300\,\text{K}$), we find an electron mobility of $\mu_{\rm{electrons}}\simeq 35{,}000\,\text{cm}^2\text{/Vs}$ and a hole mobility $\mu_{\rm{holes}}\simeq 20{,}000\,\text{cm}^2\text{/Vs}$; and, at low temperature, $\mu_{\rm{electrons}}\simeq 90{,}000\,\text{cm}^2\text{/Vs}$ and $\mu_{\rm{holes}}\simeq 30{,}000\,\text{cm}^2\text{/Vs}$.
This procedure is detailed in the SM \cite{SM}.

Moreover, it is well known that transport at low carrier density is dominated by charge puddles and thermal activation.
\cite{Martin2007, Mayorov2012, PhysRevB.84.115442}.
In order to properly extract relevant parameters for carrier scattering, we chose to avoid the regions where both electrons and holes contribute to transport, via thermal activation or potential fluctuations.
Assuming a doping independent mobility near the CNP, conductivity is given by $\sigma=ne\mu$ (magenta line in Fig.\ref{fig:figure2a}\textcolor{blue}{(d)}) and the residual carrier density tends to $n_0$ with $\sigma = n_0e\mu$ at the CNP.
Following this procedure, the residual carrier density ranges from $n_0\simeq 0.8\times 10^{11}\,\text{cm}^{-2}$ at $10\,\text{K}$ to $2.08 \times 10^{11}\,\text{cm}^{-2}$ at $300\,\text{K}$.
Those values give the correct order of magnitude when used in the phenomenological model $\sigma = \sqrt{n_0^2+n_{\rm g}^2} e \mu$ (green line in Fig. \ref{fig:figure2a}\textcolor{blue}{(d)}). Finally, the whole range of data can be reproduced properly at low doping by taking into account a gaussian disorder with potential fluctuation of $35 \,\text{meV}$ (corresponding to puddles with $n_0(0\,\text{K}) \simeq 0.74 \times 10^{11}\,\text{cm}^{-2}$) and thermal activation \cite{PhysRevB.84.115442}.
Thermal broadening contributes to the rest of the electron-hole coexistence near the Dirac point.

At this step, it is important to note that, even if doping inhomogeneity is on the order of $10^{11}\,\text{cm}^{-2}$, the actual conductivity matches the one-carrier model only above $10^{12}\,\text{cm}^{-2}$.
Moreover, in this low-density regime, extrinsic effects such as inhomogeneous strain and trapped charges can mask the intrinsic phonon contribution.
These preliminary characterizations demonstrate that we have a homogeneous, good quality graphene Hall bar suitable to study the intrinsic limits of transport as long as only carrier densities $|n| > 10^{12}\,\text{cm}^{-2}$  are considered (which corresponds roughly to Fermi energies $\varepsilon_{\rm{F}} > 100\,\text{meV}$). At higher carrier densities, enhanced screening reduces the influence of disorder, allowing acoustic and optical phonon scattering to be analyzed more reliably \cite{Efetov2010}.
For this energy and doping range, electron-phonon scattering parameters can be extracted from resistivity and compared directly with theoretical models in the next section.

\begin{figure}
    \centering
    \includegraphics[width=\linewidth]{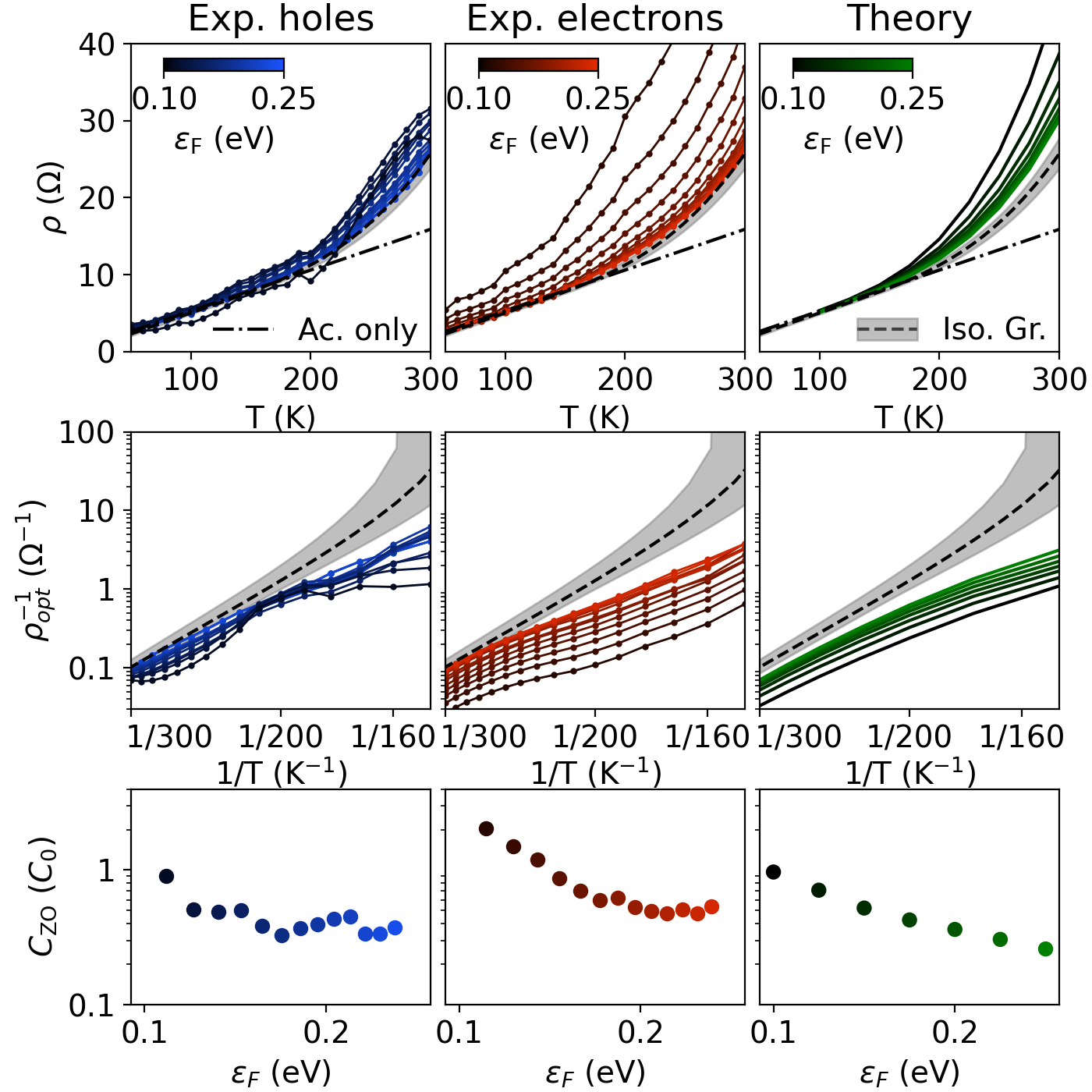}
    \caption{
    Analysis of the resistivity of BN-encapsulated graphene.
    The columns correspond to experimental measurements with hole doping (`Exp. holes')
    ,  electron doping (`Exp. electrons')
    and ab initio calculations (`Theory') considering 20 BN layers on each side of graphene.
    Each column has its own color scale. Calculations are performed for electron doping, but due to enforced electron-hole symmetry the theory compares to both experiments.
    First row shows $\rho$ as a function of $T$. The contribution from acoustic phonons alone is represented as a dash-dotted line (`Ac. only').
    Second row shows $\rho^{-1}_{\rm opt}= \left( \rho-\rho_0-\rho_{\rm ac}\right)^{-1}$ in semi-logarithmic scale as a function of $1/T$, see text.
    Last row shows the extracted electronic coupling to BN's ZO phonons, in units of $C_0 \approx 15.2$ eV$^{-1}$ ps$^{-1}$.
    In the first 2 rows, we also plot the resistivity for isolated graphene as a dashed line (`Iso. Gr.'). The grey shading represents its variations with doping level accounting also for impurity-induced residual resistivity.  .
    }
    \label{fig:Resistivity}
\end{figure}

\textit{Resistivity---} The temperature-dependent resistivity $\rho$ is shown in the first row of Fig. \ref{fig:Resistivity}, for different doping values. The first two columns show the experimental resistivities for holes and electrons extracted from the same data as in Fig. \ref{fig:figure2a}\textcolor{blue}{(c)}.

While electron-hole symmetry is broken -- as is often the case\cite{Chen2008,Farmer2009,PhysRevB.78.121402}-- both carriers display very similar resistivity trends as a function of temperature and doping magnitude, with a systematic increase upon reduction of the free carrier density. Third column  shows the results of our theoretical calculations for graphene encapsulated by twenty layers of BN on each side, where electron-hole symmetry is enforced for graphene's band structure. Twenty BN layers are sufficient to saturate the resistivity as a function of layer number (see below), and to be representative of experimental conditions where the BN thickness is estimated to be $\sim 100$ layers. The dashed black line represents the resistivity computed for isolated graphene, and the gray shading is its variation as a function of doping if only impurity scatterers are considered, as explained in the SM \cite{SM}. The dash-dotted line is instead the result obtained considering only scattering from acoustic phonons. From these results, it is evident that intrinsic scattering from isolated graphene's phonons is insufficient to explain the resistivity measurements. To support this conclusion, we now decompose the resistivity into its constituent contributions and detail their microscopic mechanisms.

\textit{Intrinsic contributions of isolated graphene---} For isolated graphene it is sufficient to solve the electronic Boltzmann transport equation (BTE) to determine the resistivity \cite{Sohier2014}. The necessary quantities entering the BTE, i.e., the Dirac cone band structure, the phonon frequencies and electron-phonon coupling matrix elements, are scaled to match GW calculations following Ref. \onlinecite{Sohier2014}. Up to 250 K, acoustics phonons strongly influence the resistivity with a linear contribution $\rho_{\rm{ac}}(T) = \alpha T$, where $\alpha$ depends on the strength of the coupling with electrons, on the mass density of graphene and the velocity of acoustic phonons, but not on the Fermi velocity. Ab initio calculations at the GW level yield a doping-independent value of $\alpha \approx 0.078$ $\Omega/K$. Linear regression of the experimental data of Fig. \ref{fig:Resistivity} between $50$ and $150$ K yields a smaller value of $\alpha \approx 0.053$ $\Omega/K$, which becomes independent of the carrier density when far-enough from the Dirac point, as detailed in the End Matter.  In our calculations, we adjusted the acoustic coupling to electrons applying a $\sim 17\%$ decrease (slope depends quadratically on the coupling) to yield $\alpha$ in agreement with measurements. At around 250 K, intrinsic optical phonons start being significantly occupied and impact the resistivity with a gradual and noticeable increase. The threshold temperature of this signature is a good indicator of the energy of the phonons involved, while the strength of the increase is related to the coupling strength with electrons.
Due to graphene's in-plane mirror symmetry, the ZO phonons of graphene do not couple.
Instead, ab initio calculations show that both optical phonons around $\Gamma$ ($\omega^{\rm Gr}_{\rm{LO/TO}} \simeq 1600$ cm$^{-1}$) and K ($\omega^{\rm Gr}_{A'_1} \simeq 1250$ cm$^{-1}$ ) contribute relevantly, due to their strong coupling that counteract the low thermal occupation \cite{Park2014,Sohier2014}.
The coupling between the $\omega_{A'_1}$ phonon and electrons is particularly strong due to electron-electron interactions \cite{Guandalini2025,Caldarelli2025}, and sensitive to doping and environment conditions. For this reason it is much more difficult to determine accurately with respect to the other couplings. In the SM \cite{SM}, we show that the $\omega_{A'_1}$ phonon frequency itself is sensitive to doping and environment conditions, but it stays high enough in energy to be excluded as a relevant contribution to resistivity below 250 K even at low doping, and the single value $\omega^{\rm Gr}_{A'_1} \simeq 1250$ cm$^{-1}$ is a reasonable approximation across the studied experimental conditions. Therefore, intrinsic graphene mechanisms cannot explain the exponential increase of the experimental resistivity witnessed around 150 K in BN-encapsulated graphene. Given the intrinsically weak scattering from graphene's intrinsic phonons at low temperatures, which yields high mobilities in high-quality devices, transport can then be limited by contributions that are typically negligible in lower-mobility materials, namely BN's phonons.

\textit{Role of BN phonons---}
To study the scattering from BN's phonons, we solve the coupled dynamical Boltzmann equations for the carriers and the electrodynamic modes, in order to account for the subtle interplay of phonon and electron-hole pair excitations. We notably include the contribution of electron-electron scattering and the dynamical screening of phonons, which is essential to get correct results \cite{accpaper}.

Several calculations, reported in the SM \cite{SM}, were conducted to establish the following results.
Both in-plane ($\omega^{\rm BN}_{\rm LO} \sim 1500 $ cm$^{-1}$) and out-of-plane ($\omega^{\rm BN}_{\rm ZO} \sim 800$ cm$^{-1}$) polar-optical phonons in BN couple remotely to electrons in graphene. In both cases, it is important to highlight that they are not a single remote interfacial phonon (RIP) \cite{PhysRevB.6.4517,HESS1979797,PhysRevB.86.165422}, but they are simply the $N$ LO and ZO modes of $N$-layers BN. As for all phonon modes, the scattering of graphene's electrons by LO and ZO depends both on the coupling strength and the phonon's populations. As shown in the SM \cite{SM}, LO phonons display a stronger coupling in general, but they are more energetic so less populated, and they become important at higher temperatures, mostly above room temperature. ZO phonons couple less strongly, but they are half as energetic as the LO phonons, which means they are much more populated ($\sim 1000$ times more than LO at 150K).
Interestingly, as the number of BN layers increase, the contribution to resistivity from LO phonons decreases, while that of ZO increases until both saturate after a dozen layers. This trend is hard to detect with standard linear response ab initio techniques that are computationally very expensive for many-layer materials.
Overall, below room temperature and for more than a dozen BN layers, ZO phonons clearly dominate the resistivity due to remote scattering.
As anticipated, calculations including the effects of BN phonons recreate the additional contribution to resistivity seen in experiments. A clearer comparison is performed by plotting $\rho^{-1}_{\rm opt}=\left( \rho-\rho_0-\rho_{\rm ac}\right)^{-1} $
in a semi-logarithmic scale versus $1/T$, as done in the second row of Fig. \ref{fig:Resistivity} applying a three-neighbour smoothing to suppress discretization artifacts. $\rho_0$ is the (small) estimated residual resistivity, obtained from linear regression in the linear regime for resistivity, and $\rho_{\rm ac}$ is the resistivity due to acoustic phonon only. We choose this quantity because, assuming additive contributions to resistivity from different scattering mechanisms, $\rho_{\rm opt}$ becomes proportional to the contribution from optical phonons. Then, if we further assume that $\rho_{\rm opt}$ is roughly proportional to the Bose-Einstein distribution occupation of the corresponding optical phonons, i.e. $\rho_{\rm opt} \propto n(\omega_{\rm ph})$, and that $e^{\hbar \omega_{\rm ph}/kT} \gg 1$ (reasonable since $\hbar \omega_{\rm ph}> 3 k_{\rm B}T$ ), then $\rho^{-1}_{\rm opt} \propto e^{\hbar \omega_{\rm ph}/kT}$. Therefore, experimental data of the second row should show linear curves with a slope determined by the phonon frequency, if the previous assumptions are accurate. In practice, this is not exactly the case (see discussion about the additivity of resistivities in SM \cite{SM}), and we cannot quantitatively estimate the phonon frequency involved in resistivity just by looking at its slope. Nonetheless, $\rho^{-1}_{\rm opt}$ still remains a qualitatively useful quantity to discern between strongly different scattering mechanisms. Indeed, both experiments and theoretical calculations display curves with similar slopes in the case of BN-encapsulated graphene, that are the marker of scattering from remote ZO phonons, while the curve corresponding to simulations for isolated graphene clearly displays different slope and magnitude.

Finally, we extract the experimental coupling between electrons and BN's ZO phonons with a procedure described in SM \cite{SM}, and compare it to the coupling obtained from theoretical curves using the same extraction procedure. In short, we set up an electronic BTE including acoustic phonons scattering, with couplings fitted on experiments, and two optical phonons whose coupling strengths with electrons are treated as adjustable parameters. The frequencies of those two optical phonons are fixed: one to BN's ZO phonon $\omega^{\rm BN}_{\rm ZO} = 800$ cm$^{-1}$, the other to the lowest graphene's optical phonon $\omega^{\rm Gr}_{A'_1}=1225$ cm$^{-1}$. The extraction procedure for the isolated graphene simulations reassuringly yield zero coupling to BN's ZO phonons. The coupling strength obtained when dealing with BN-encapsulated graphene are presented in the third row of Fig. \ref{fig:Resistivity}. They are of similar magnitude in experiments and theory. Importantly, the doping dependency is very well reproduced: at lower dopings the coupling with phonons is enhanced due to reduced screening.

\textit{Conclusions---}
In summary we find that, between 150 K and room temperature, the temperature-dependant resistivity of graphene is strongly influenced by the coupling to hBN's out-of-plane ZO phonons, whereas in-plane LO modes play a negligible role. This behaviour is highly sensitive to the hBN thickness: the ZO contribution grows with the number of layers, while the LO contribution diminishes, reaching a saturation after around a dozen layers on each side, a trend that cannot be captured by simple interfacial phonon models.
Crucially, both experiments and theory conclude that remote phonon scattering becomes increasingly important at lower doping levels. We attribute this to the reduced screening efficiency of graphene as the free carrier density decreases.
These results demonstrate that, even in high-quality devices, the conductivity of hBN-encapsulated graphene is ultimately limited by the coupling to hBN phonons, establishing a fundamental extrinsic limit to charge transport.
Thus, the intrinsic transport performances of graphene below and up to room temperature cannot truly be reached with BN encapsulation, although they are certainly approached at large doping.% ($n>3 \ 10^{12}$ cm$^{-2}$).

\textit{Acknowledgments---}
We acknowledge the Institute for Quantum Technologies in Occitanie (IQO) and the Occitanie Region for their support through a doctoral fellowship. Additional funding was provided by the French Agence Nationale de la Recherche (ANR) under the Jjedi project
(JCJC, ANR-18-CE24-0004) and the VanaSiC project (PRC, ANR-24-CE24-0022).
We acknowledge the EuroHPC Joint Undertaking for awarding this project access to the EuroHPC supercomputer LUMI, hosted by CSC (Finland) and the LUMI consortium through a EuroHPC Regular Access call.
This work has been realized with the support of the ISDM-MESO Platform at the University of Montpellier funded under the CPER by the French Government, the Occitanie Region, the Metropole of Montpellier and the University of Montpellier.

\clearpage

\appendix
\onecolumngrid
\begin{center}
\textbf{END MATTER}
\end{center}
\twocolumngrid

\begin{figure}
    \centering
    \includegraphics[width=0.98\linewidth]{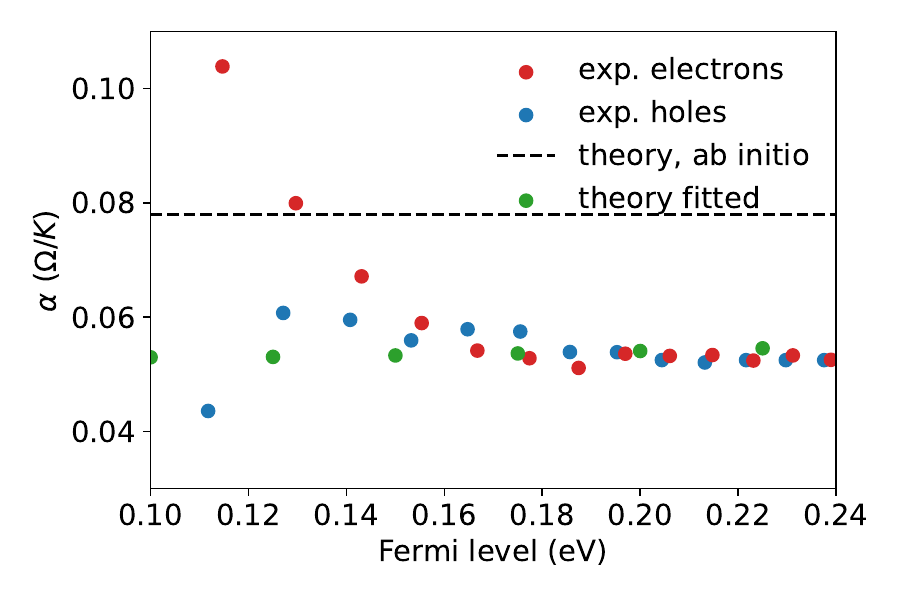}
    \caption{Slope of the acoustic phonons' linear contribution to the resistivity as a function of Fermi level. The slope is extracted from experiments by linear regression on the resistivity between $50$ and $150$ K, for both electron (`electrons' --- red circles)  and hole (`holes' --- blue circles) doping type. The result from ab initio calculations, performed at the GW level, is represented as a constant line (`theory, ab initio' --- dashed line). The same fitting procedure is then applied to the theoretical resistivity calculations, where the acoustic coupling is reduced by $17\%$  (`theory fitted' --- green dots). This consistently shows good agreement with experimental data in a significant range of Fermi levels.}
    \label{fig:AcSlope}
\end{figure}

\label{app:endmatter}
\textit{Appendix A: Acoustic phonon contribution}--- The linear behavior of graphene's resistivity up to $150$ K has been systematically attributed to acoustic phonons. It is well reproduced by ab initio simulations, although the value of the associated slope can vary between different experiments. In most experiments where this parameter was fitted, whether in supported graphene \cite{Chen2008,Efetov2010} or BN encapsulated graphene \cite{Wang2013a,Banszerus2019}, the slope was found higher than predicted by ab initio calculations, with $\alpha \approx 0.1$ $\Omega/K$. This value corresponds to a $15\%$ increase with respect to the \textit{ab initio} coupling between electrons and acoustic phonons \cite{Sohier2014}.
Fig. \ref{fig:AcSlope} shows the slopes extracted from the resistivities measured in this work, for both electron and hole doping. They are significantly lower than previous works, although they also correspond to a relatively mild deviation from the \textit{ab initio} coupling to acoustic phonons, i.e. a $17\%$ decrease. While intriguing, we leave the investigation of this aspect for future works.

\textit{Appendix B: Measurement details and carrier density assessment---}
Using standard dry transfer techniques \cite{Frisenda2018}, a graphene monolayer is placed between two layers of $\sim 30\,\text{nm}$ thick h-BN, and transferred on top of a (n$^+$ doped) Si/$300\,\text{nm}$ SiO$_2$ substrate.
The resulting device is patterned by e-beam lithography into a Hall bar geometry with a channel length of $2\,\mu\text{m}$, a width of $4\,\mu\text{m}$ between lateral contacts and etched using CHF$_3$ plasma etching in a Corial ICP reactive ion etching system.
A final lithography step is done to deposit gold quasi-1D edge contacts by lift-off.

Transport measurements were carried out in a Janis closed-cycle helium cryostat in the 10 to 300 K range, combined with a 1 T electromagnet. Transverse and longitudinal conductivities were obtained by injecting a 100 nA current at 33.33 Hz between source and drain (S and D in figure \ref{fig:figure2a}.a) and using lock-ins between 1-4 and 4-3 respectively. A backgate voltage, $V_g$, was applied between the Si and the drain with a source measure unit (allowing the monitoring of any undesired leakage current), typically in the range $[-50;100]\,\textrm{V}$. Gate-dependent Hall measurements were performed at fixed and low enough magnetic fields to avoid nonlinearities or quantum effects, and the linearity of $R_{\rm{H}}$ was also regularly checked at fixed $V_{\rm{g}}$ by sweeping the magnetic field.
\begin{figure}
    \centering
    \includegraphics[width=0.98\linewidth]{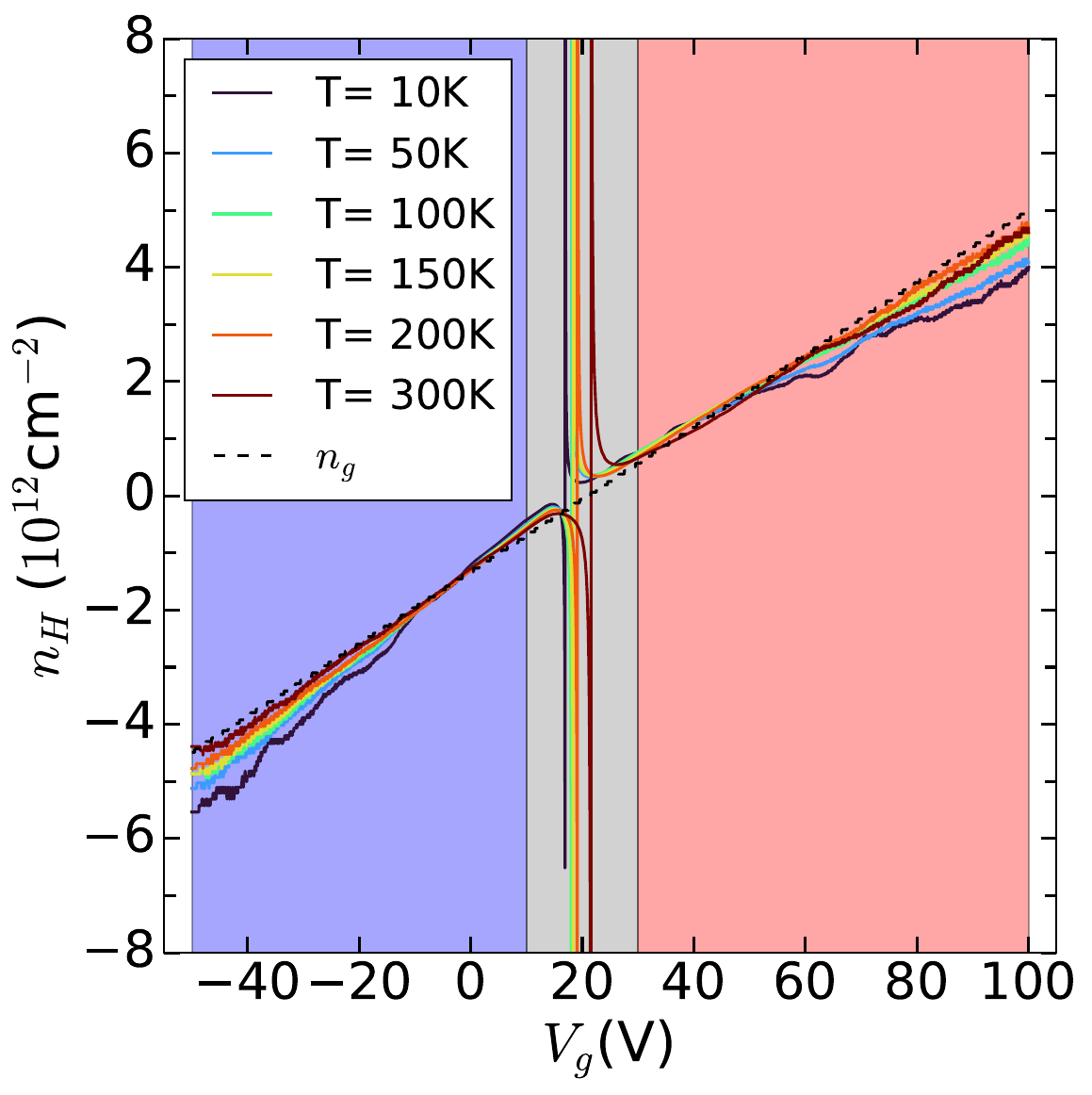}
    \caption{Gate dependent carrier density, $n_{\textrm{H}}$, determined using the Hall voltage every 50 K, and $n_{\textrm{g}}$ calculated with a capacitive model with $C_{\textrm{g}} = 10^{-4}$ F.m$^{-2}$. The gray area is where $n_{\rm{H}} \neq n_{\rm{g}}$, while the blue and red ones are for regions where it can be safely stated that only one type of carrier contributes to transport.}
    \label{fig:ng_nh_exp}
\end{figure}
Classical Hall effect was used to determine the carrier density $n_{\rm H}(V_{\rm{g}})$ every 50 K which was consistent with a bottom hBN thickness of $30\,\text{nm}$ obtained from a plane capacitor model, as shown in Fig. \ref{fig:ng_nh_exp}. Longitudinal conductivity was measured every 10 K and could be plotted as a function of $\Delta V_{\rm{g}} = (V_{\rm{g}}-V_{\rm{CNP}})$, where $V_{\rm{CNP}}$ corresponds to the gate voltage at minimum conductivity and the charge neutrality point of graphene (Fig. \ref{fig:figure2a}.c). The position of the CNP obtained with this method and using $n_{\rm{H}} = 0\,\text{cm}^{-2}$ are systematically consistent. Moreover, the gray area in Fig. \ref{fig:ng_nh_exp} where $n_{\rm{H}} \neq n_{\rm{g}}$, correspond to the same values of disorder as the one discussed in the main text, while the blue and red ones are for $\varepsilon_{\rm{F}} \ge 100\, \text{meV}$. Mobilities extracted from this different methods are also consistent (see SM \cite{SM}).

Supplemental Material includes Refs. \cite{Trolle2017,Guandalini2025a,Carnimeo2023,Hellman2013,Martin_Reining_Ceperley_2016,PhysRevB.82.165111,PhysRevB.111.024307,PhysRevX.13.041009,PhysRevLett.68.3603,PhysRevLett.92.075501,PhysRevB.76.035439,PhysRevB.80.085423,PhysRevLett.93.185503,jsan2010085,Macheda2020,Perdew_1996,Binci_2021,Laturia_2018}

\bibliography{VEDcBTE_BNGr_ZO}

@article{Frisenda2018,
  title = {Recent Progress in the Assembly of Nanodevices and van Der {{Waals}} Heterostructures by Deterministic Placement of {{2D}} Materials},
  author = {Frisenda, Riccardo and {Navarro-Moratalla}, Efr{\'e}n and Gant, Patricia and P{\'e}rez De Lara, David and {Jarillo-Herrero}, Pablo and Gorbachev, Roman V. and {Castellanos-Gomez}, Andres},
  year = 2018,
  journal = {Chemical Society Reviews},
  volume = {47},
  number = {1},
  pages = {53--68},
  issn = {0306-0012, 1460-4744},
  doi = {10.1039/C7CS00556C},
  langid = {english}
}

@article{Banszerus2019,
  title = {Extraordinary High Room-Temperature Carrier Mobility in Graphene-{{WSe}}{$_2$} Heterostructures},
  author = {Banszerus, L. and Sohier, T. and Epping, A. and Winkler, F. and Libisch, F. and Haupt, F. and Watanabe, K. and Taniguchi, T. and {M{\"u}ller-Caspary}, K. and Marzari, N. and Mauri, F. and Beschoten, B. and Stampfer, C.},
  year = 2019,
  month = sep,
  journal = {ArXiv:1909.09523},
  eprint = {1909.09523},
  archiveprefix = {arXiv},
  arxivid = {1909.09523},
  copyright = {All rights reserved}
}

@article{Baudin2020,
  title = {Hyperbolic {{Phonon Polariton Electroluminescence}} as an {{Electronic Cooling Pathway}}},
  author = {Baudin, Emmanuel and Voisin, Christophe and Pla{\c c}ais, Bernard},
  year = 2020,
  month = feb,
  journal = {Advanced Functional Materials},
  volume = {30},
  number = {8},
  pages = {1904783},
  issn = {1616-301X, 1616-3028},
  doi = {10.1002/adfm.201904783},
  langid = {english}
}

@article{Caldarelli2025,
  title = {Variational Formulation of Dynamical Electronic Response Functions in the Presence of Nonlocal Exchange Interactions},
  author = {Caldarelli, Giovanni and Guandalini, Alberto and Macheda, Francesco and Mauri, Francesco},
  year = 2025,
  month = feb,
  journal = {Physical Review B},
  volume = {111},
  number = {7},
  pages = {075137},
  publisher = {American Physical Society},
  doi = {10.1103/PhysRevB.111.075137}
}

@article{Carnimeo2023,
  title = {Quantum {{ESPRESSO}}: {{One}} Further Step toward the Exascale},
  author = {Carnimeo, Ivan and Affinito, Fabio and Baroni, Stefano and Baseggio, Oscar and Bellentani, Laura and Bertossa, Riccardo and Delugas, Pietro Davide and Ruffino, Fabrizio Ferrari and Orlandini, Sergio and Spiga, Filippo and Giannozzi, Paolo},
  year = 2023,
  month = oct,
  journal = {Journal of Chemical Theory and Computation},
  volume = {19},
  number = {20},
  pages = {6992--7006},
  publisher = {American Chemical Society},
  issn = {1549-9618},
  doi = {10.1021/acs.jctc.3c00249}
}

@article{Chen2008,
  ids = {Chen2008a,Chen2008b},
  title = {Intrinsic and Extrinsic Performance Limits of Graphene Devices on {{SiO2}}},
  author = {Chen, JH Jian-Hao and Jang, Chaun and Xiao, Shudong and Ishigami, Masa and Fuhrer, Michael S},
  year = 2008,
  month = apr,
  journal = {Nature Nanotechnology},
  volume = {3},
  number = {April},
  eprint = {18654504},
  eprinttype = {pubmed},
  pages = {1--4},
  publisher = {Nature Publishing Group},
  issn = {1748-3395},
  doi = {10.1038/nnano.2008.58},
  copyright = {2008 Springer Nature Limited},
  pmid = {18654504}
}

@article{Chen2022,
  title = {Remote Surface Optical Phonon Scattering in Ferroelectric {{Ba0}}.{{6Sr0}}.{{4TiO3}} Gated Graphene},
  author = {Chen, Hanying and Li, Tianlin and Hao, Yifei and Rajapitamahuni, Anil and Xiao, Zhiyong and Schoeche, Stefan and Schubert, Mathias and Hong, Xia},
  year = 2022,
  month = oct,
  journal = {Journal of Applied Physics},
  volume = {132},
  number = {15},
  pages = {154301},
  issn = {0021-8979},
  doi = {10.1063/5.0106939}
}

@article{DasSarma2011,
  ids = {DasSarma2011a},
  title = {Electronic Transport in Two-Dimensional Graphene},
  author = {Das Sarma, S. and Adam, Shaffique and Hwang, E. H. and Rossi, Enrico and Sarma, S Das and Adam, Shaffique and Rossi, Enrico and Das Sarma, S. and Adam, Shaffique and Hwang, E. H. and Rossi, Enrico},
  year = 2011,
  month = may,
  journal = {Reviews of Modern Physics},
  volume = {83},
  number = {2},
  pages = {407--470},
  publisher = {American Physical Society},
  issn = {0034-6861},
  doi = {10.1103/RevModPhys.83.407}
}

@article{Dean2010,
  ids = {Dean2010a},
  title = {Boron Nitride Substrates for High-Quality Graphene Electronics.},
  shorttitle = {Nat Nano},
  author = {Dean, C R and Young, A F and Meric, I and Lee, C and Wang, L and Sorgenfrei, S and Watanabe, K and Taniguchi, T and Kim, P and Shepard, K L and Hone, J},
  year = 2010,
  month = oct,
  journal = {Nature nanotechnology},
  volume = {5},
  number = {10},
  eprint = {20729834},
  eprinttype = {pubmed},
  pages = {722--6},
  publisher = {Nature Publishing Group},
  issn = {1748-3395},
  doi = {10.1038/nnano.2010.172},
  copyright = {2010 Springer Nature Limited},
  pmid = {20729834}
}

@article{Efetov2010,
  ids = {Efetov2010a},
  title = {Controlling Electron-Phonon Interactions in Graphene at Ultrahigh Carrier Densities},
  author = {Efetov, Dmitri K. and Kim, Philip},
  year = 2010,
  month = dec,
  journal = {Physical Review Letters},
  volume = {105},
  number = {25},
  pages = {256805},
  issn = {0031-9007},
  doi = {10.1103/PhysRevLett.105.256805}
}

@article{Fratini2008,
  ids = {Fratini2008a},
  title = {Substrate-Limited Electron Dynamics in Graphene},
  author = {Fratini, S. and Guinea, F.},
  year = 2008,
  month = may,
  journal = {Physical Review B},
  volume = {77},
  number = {19},
  eprint = {0711.1303v3},
  pages = {195415},
  publisher = {American Physical Society},
  issn = {1098-0121},
  doi = {10.1103/PhysRevB.77.195415},
  archiveprefix = {arXiv},
  arxivid = {arXiv:0711.1303v3}
}

@article{Guandalini2025,
  title = {Excitonic {{Effects}} in {{Phonons}}: {{Reshaping}} the {{Graphene Kohn Anomalies}} and {{Lifetimes}}},
  shorttitle = {Excitonic {{Effects}} in {{Phonons}}},
  author = {Guandalini, Alberto and Macheda, Francesco and Caldarelli, Giovanni and Mauri, Francesco},
  year = 2025,
  month = aug,
  journal = {Physical Review Letters},
  volume = {135},
  number = {7},
  pages = {076401},
  issn = {0031-9007, 1079-7114},
  doi = {10.1103/y1dn-m6pc},
  langid = {english}
}

@article{Guandalini2025a,
  title = {High- and Low-Energy Many-Body Effects of Graphene in a Unified Approach},
  author = {Guandalini, Alberto and Caldarelli, Giovanni and Macheda, Francesco and Mauri, Francesco},
  year = 2025,
  month = feb,
  journal = {Physical Review B},
  volume = {111},
  number = {7},
  pages = {075118},
  publisher = {American Physical Society},
  doi = {10.1103/PhysRevB.111.075118}
}

@article{Guo2025a,
  title = {Hyperbolic Phonon-Polariton Electroluminescence in {{2D}} Heterostructures},
  author = {Guo, Qiushi and Esin, Iliya and Li, Cheng and Chen, Chen and Han, Guanyu and Liu, Song and Edgar, James H. and Zhou, Selina and Demler, Eugene and Refael, Gil and Xia, Fengnian},
  year = 2025,
  month = mar,
  journal = {Nature},
  volume = {639},
  number = {8056},
  pages = {915--921},
  issn = {0028-0836, 1476-4687},
  doi = {10.1038/s41586-025-08686-9},
  langid = {english}
}

@article{Hellman2013,
  title = {Temperature Dependent Effective Potential Method for Accurate Free Energy Calculations of Solids},
  author = {Hellman, Olle and Steneteg, Peter and Abrikosov, I. A. and Simak, S. I.},
  year = 2013,
  month = mar,
  journal = {Physical Review B},
  volume = {87},
  number = {10},
  pages = {104111},
  publisher = {American Physical Society},
  doi = {10.1103/PhysRevB.87.104111}
}

@article{Hwang2008,
  ids = {Hwang2008b},
  title = {Acoustic Phonon Scattering Limited Carrier Mobility in Two-Dimensional Extrinsic Graphene},
  author = {Hwang, E. H. and Das Sarma, S.},
  year = 2008,
  journal = {Physical Review B},
  volume = {77},
  number = {11},
  pages = {115449},
  publisher = {American Physical Society}
}

@article{Kaasbjerg2012,
  title = {Unraveling the Acoustic Electron-Phonon Interaction in Graphene},
  shorttitle = {Phys. {{Rev}}. {{B}}},
  author = {Kaasbjerg, Kristen and Thygesen, Kristian S. KS Kristian and Jacobsen, KW Karsten W.},
  year = 2012,
  month = apr,
  journal = {Physical Review B},
  volume = {85},
  number = {16},
  eprint = {1201.4661v1},
  pages = {23--27},
  publisher = {American Physical Society},
  issn = {1098-0121},
  doi = {10.1103/PhysRevB.85.165440},
  archiveprefix = {arXiv},
  arxivid = {arXiv:1201.4661v1}
}

@article{Konar2010,
  title = {Effect of High- {$\kappa$} Gate Dielectrics on Charge Transport in Graphene-Based Field Effect Transistors},
  author = {Konar, Aniruddha and Fang, Tian and Jena, Debdeep},
  year = 2010,
  month = sep,
  journal = {Physical Review B},
  volume = {82},
  number = {11},
  pages = {115452},
  issn = {1098-0121, 1550-235X},
  doi = {10.1103/PhysRevB.82.115452},
  copyright = {http://link.aps.org/licenses/aps-default-license},
  langid = {english}
}

@article{Lin2013,
  title = {Surface Polar Optical Phonon Scattering of Carriers in Graphene on Various Substrates},
  author = {Lin, I-Tan and Liu, Jia-Ming},
  year = 2013,
  month = aug,
  journal = {Applied Physics Letters},
  volume = {103},
  number = {8},
  pages = {081606},
  issn = {0003-6951},
  doi = {10.1063/1.4819395}
}

@article{Macheda2020,
  ids = {Macheda2020a},
  title = {Theory and {{Computation}} of {{Hall Scattering Factor}} in {{Graphene}}},
  author = {Macheda, Francesco and Ponc{\'e}, Samuel and Giustino, Feliciano and Bonini, Nicola},
  year = 2020,
  month = dec,
  journal = {Nano Letters},
  volume = {20},
  number = {12},
  pages = {8861--8865},
  publisher = {American Chemical Society},
  issn = {1530-6984, 1530-6992},
  doi = {10.1021/acs.nanolett.0c03874},
  langid = {english}
}

@article{Macheda2024b,
  title = {{\emph{Ab Initio}} {{Van}} Der {{Waals}} Electrodynamics: {{Polaritons}} and Electron Scattering from Plasmons and Phonons in {{BN-capped}} Graphene},
  shorttitle = {{\emph{Ab Initio}} {{Van}} Der {{Waals}} Electrodynamics},
  author = {Macheda, Francesco and Mauri, Francesco and Sohier, Thibault},
  year = 2024,
  month = sep,
  journal = {Physical Review B},
  volume = {110},
  number = {11},
  pages = {115407},
  issn = {2469-9950, 2469-9969},
  doi = {10.1103/PhysRevB.110.115407},
  copyright = {All rights reserved},
  langid = {english}
}

@article{Martin2007,
  ids = {Martin2008},
  title = {Observation of Electron--Hole Puddles in Graphene Using a Scanning Single-Electron Transistor},
  shorttitle = {Nat Phys},
  author = {Martin, J. and Akerman, N. and Ulbricht, G. and Lohmann, T. and Smet, J. H. and Von Klitzing, K. and Yacoby, A.},
  year = 2007,
  month = nov,
  journal = {Nature Physics},
  volume = {4},
  number = {2},
  pages = {144--148},
  publisher = {Nature Publishing Group},
  issn = {1745-2473},
  doi = {10.1038/nphys781},
  copyright = {2008 Springer Nature Limited}
}

@article{Mayorov2012,
  title = {How {{Close Can One Approach}} the {{Dirac Point}} in {{Graphene Experimentally}}?},
  author = {Mayorov, Alexander S. and Elias, Daniel C. and Mukhin, Ivan S. and Morozov, Sergey V. and Ponomarenko, Leonid A. and Novoselov, Kostya S. and Geim, A. K. and Gorbachev, Roman V.},
  year = 2012,
  month = sep,
  journal = {Nano Letters},
  volume = {12},
  number = {9},
  pages = {4629--4634},
  publisher = {American Chemical Society},
  issn = {1530-6984},
  doi = {10.1021/nl301922d}
}

@article{Ong2012,
  title = {Theory of Interfacial Plasmon-Phonon Scattering in Supported Graphene},
  author = {Ong, Zhun-Yong and Fischetti, Massimo V.},
  year = 2012,
  month = oct,
  journal = {Physical Review B},
  volume = {86},
  number = {16},
  pages = {165422},
  issn = {1098-0121, 1550-235X},
  doi = {10.1103/PhysRevB.86.165422},
  langid = {english}
}

@article{Park2014,
  ids = {Park2014a},
  title = {Electron-Phonon Interactions and the Intrinsic Electrical Resistivity of Graphene.},
  author = {Park, Cheol-Hwan and Bonini, Nicola and Sohier, Thibault and Samsonidze, Georgy and Kozinsky, Boris and Calandra, Matteo and Mauri, Francesco and Marzari, Nicola},
  year = 2014,
  month = feb,
  journal = {Nano letters},
  volume = {14},
  number = {3},
  pages = {1113--1119},
  publisher = {American Chemical Society},
  issn = {1530-6992},
  doi = {10.1021/nl402696q},
  copyright = {All rights reserved},
  pmid = {24524418}
}

@article{Perebeinos2010,
  ids = {Perebeinos2010a},
  title = {Inelastic Scattering and Current Saturation in Graphene},
  author = {Perebeinos, Vasili and Avouris, Phaedon},
  year = 2010,
  month = may,
  journal = {Physical Review B},
  volume = {81},
  number = {19},
  pages = {195442},
  issn = {1098-0121},
  doi = {10.1103/PhysRevB.81.195442}
}

@article{Schiefele2012,
  ids = {Schiefele2012a},
  title = {Temperature Dependence of the Conductivity of Graphene on Boron Nitride},
  author = {Schiefele, J{\"u}rgen and Sols, Fernando and Guinea, Francisco},
  year = 2012,
  month = may,
  journal = {Physical Review B},
  volume = {85},
  number = {19},
  eprint = {1202.2440},
  primaryclass = {cond-mat},
  pages = {195420},
  issn = {1098-0121},
  doi = {10.1103/PhysRevB.85.195420},
  archiveprefix = {arXiv}
}

@article{Sohier2014,
  ids = {Sohier2014a,Sohier_2014,Sohier₂014},
  title = {Phonon-Limited Resistivity of Graphene by First-Principles Calculations: {{Electron-phonon}} Interactions, Strain-Induced Gauge Field, and {{Boltzmann}} Equation},
  author = {Sohier, Thibault and Calandra, Matteo and Park, C.-H. Cheol-Hwan and Bonini, Nicola and Marzari, Nicola and Mauri, Francesco},
  year = 2014,
  month = sep,
  journal = {Physical Review B},
  volume = {90},
  number = {12},
  pages = {125414},
  issn = {1098-0121},
  doi = {10.1103/PhysRevB.90.125414}
}

@article{Trolle2017,
  title = {Model Dielectric Function for {{2D}} Semiconductors Including Substrate Screening},
  author = {Trolle, Mads L. and Pedersen, Thomas G. and V{\'e}niard, Valerie},
  year = 2017,
  month = jan,
  journal = {Scientific Reports},
  volume = {7},
  pages = {39844},
  publisher = {Nature Publishing Group},
  issn = {2045-2322},
  doi = {10.1038/srep39844}
}

@article{Venanzi2023,
  ids = {Venanzi2023a},
  title = {Probing {{Enhanced Electron-Phonon Coupling}} in {{Graphene}} by {{Infrared Resonance Raman Spectroscopy}}},
  author = {Venanzi, Tommaso and Graziotto, Lorenzo and Macheda, Francesco and Sotgiu, Simone and Ouaj, Taoufiq and Stellino, Elena and Fasolato, Claudia and Postorino, Paolo and Mi{\v s}eikis, Vaidotas and Metzelaars, Marvin and K{\"o}gerler, Paul and Beschoten, Bernd and Coletti, Camilla and Roddaro, Stefano and Calandra, Matteo and Ortolani, Michele and Stampfer, Christoph and Mauri, Francesco and Baldassarre, Leonetta},
  year = 2023,
  month = jun,
  journal = {Physical Review Letters},
  volume = {130},
  number = {25},
  pages = {256901},
  publisher = {American Physical Society},
  issn = {0031-9007, 1079-7114},
  doi = {10.1103/PhysRevLett.130.256901},
  langid = {english}
}

@article{Wang2013a,
  ids = {Wang2013b},
  title = {One-Dimensional Electrical Contact to a Two-Dimensional Material.},
  author = {Wang, L and Meric, I and Huang, P Y and Gao, Q and Gao, Y and Tran, H and Taniguchi, T and Watanabe, K and Campos, L M and a Muller, D and Guo, J and Kim, P and Hone, J and Shepard, K L and Dean, C R},
  year = 2013,
  month = nov,
  journal = {Science},
  volume = {342},
  number = {6158},
  eprint = {24179223},
  eprinttype = {pubmed},
  pages = {614--7},
  publisher = {American Association for the Advancement of Science},
  issn = {1095-9203},
  doi = {10.1126/science.1244358},
  pmid = {24179223}
}

@article{Xue2011,
  title = {Scanning Tunnelling Microscopy and Spectroscopy of Ultra-Flat Graphene on Hexagonal Boron Nitride},
  author = {Xue, Jiamin and {Sanchez-Yamagishi}, Javier and Bulmash, Danny and Jacquod, Philippe and Deshpande, Aparna and Watanabe, K. and Taniguchi, T. and {Jarillo-Herrero}, Pablo and LeRoy, Brian J.},
  year = 2011,
  month = apr,
  journal = {Nature Materials},
  volume = {10},
  number = {4},
  pages = {282--285},
  publisher = {Nature Publishing Group},
  issn = {1476-4660},
  doi = {10.1038/nmat2968},
  copyright = {2011 Springer Nature Limited},
  langid = {english}
}

@article{Yang2018,
  title = {A Graphene {{Zener}}--{{Klein}} Transistor Cooled by a Hyperbolic Substrate},
  author = {Yang, Wei and Berthou, Simon and Lu, Xiaobo and Wilmart, Quentin and Denis, Anne and Rosticher, Michael and Taniguchi, Takashi and Watanabe, Kenji and F{\`e}ve, Gwendal and Berroir, Jean-Marc and Zhang, Guangyu and Voisin, Christophe and Baudin, Emmanuel and Pla{\c c}ais, Bernard},
  year = 2018,
  month = jan,
  journal = {Nature Nanotechnology},
  volume = {13},
  number = {1},
  pages = {47--52},
  issn = {1748-3387, 1748-3395},
  doi = {10.1038/s41565-017-0007-9},
  langid = {english}
}

@article{You2019,
  title = {Role of Remote Interfacial Phonons in the Resistivity of Graphene},
  author = {You, Y. G. and Ahn, J. H. and Park, B. H. and Kwon, Y. and Campbell, E. E. B. and Jhang, S. H.},
  year = 2019,
  month = jul,
  journal = {Applied Physics Letters},
  volume = {115},
  number = {4},
  pages = {043104},
  issn = {0003-6951},
  doi = {10.1063/1.5097043}
}

@article{PhysRevB.96.014101,
  title = {Suppression of intrinsic roughness in encapsulated graphene},
  author = {Thomsen, Joachim Dahl and Gunst, Tue and Gregersen, S\o{}ren Schou and Gammelgaard, Lene and Jessen, Bjarke S\o{}rensen and Mackenzie, David M. A. and Watanabe, Kenji and Taniguchi, Takashi and B\o{}ggild, Peter and Booth, Timothy J.},
  journal = {Phys. Rev. B},
  volume = {96},
  issue = {1},
  pages = {014101},
  numpages = {8},
  year = {2017},
  month = {Jul},
  publisher = {American Physical Society},
  doi = {10.1103/PhysRevB.96.014101},
  url = {https://link.aps.org/doi/10.1103/PhysRevB.96.014101}
}

@Article{Farmer2009,
author={Farmer, Damon B.
and Golizadeh-Mojarad, Roksana
and Perebeinos, Vasili
and Lin, Yu-Ming
and Tulevski, George S.
and Tsang, James C.
and Avouris, Phaedon},
title={Chemical Doping and Electron−Hole Conduction Asymmetry in Graphene Devices},
journal={Nano Letters},
year={2009},
month={Jan},
day={14},
publisher={American Chemical Society},
volume={9},
number={1},
pages={388-392},
issn={1530-6984},
doi={10.1021/nl803214a},
url={https://doi.org/10.1021/nl803214a}
}

@article{PhysRevB.78.121402,
  title = {Evidence of the role of contacts on the observed electron-hole asymmetry in graphene},
  author = {Huard, B. and Stander, N. and Sulpizio, J. A. and Goldhaber-Gordon, D.},
  journal = {Phys. Rev. B},
  volume = {78},
  issue = {12},
  pages = {121402},
  numpages = {4},
  year = {2008},
  month = {Sep},
  publisher = {American Physical Society},
  doi = {10.1103/PhysRevB.78.121402},
  url = {https://link.aps.org/doi/10.1103/PhysRevB.78.121402}
}

@article{PhysRevB.6.4517,
  title = {Electron Scattering from Surface Excitations},
  author = {Wang, S. Q. and Mahan, G. D.},
  journal = {Phys. Rev. B},
  volume = {6},
  issue = {12},
  pages = {4517--4524},
  numpages = {0},
  year = {1972},
  month = {Dec},
  publisher = {American Physical Society},
  doi = {10.1103/PhysRevB.6.4517},
  url = {https://link.aps.org/doi/10.1103/PhysRevB.6.4517}
}

@article{HESS1979797,
title = {Remote polar phonon scattering in silicon inversion layers},
journal = {Solid State Communications},
volume = {30},
number = {12},
pages = {797-799},
year = {1979},
issn = {0038-1098},
doi = {https://doi.org/10.1016/0038-1098(79)90051-6},
url = {https://www.sciencedirect.com/science/article/pii/0038109879900516},
author = {K. Hess and P. Vogl}
}

@article{PhysRevB.86.165422,
  title = {Theory of interfacial plasmon-phonon scattering in supported graphene},
  author = {Ong, Zhun-Yong and Fischetti, Massimo V.},
  journal = {Phys. Rev. B},
  volume = {86},
  issue = {16},
  pages = {165422},
  numpages = {15},
  year = {2012},
  month = {Oct},
  publisher = {American Physical Society},
  doi = {10.1103/PhysRevB.86.165422},
  url = {https://link.aps.org/doi/10.1103/PhysRevB.86.165422}
}

@book{Martin_Reining_Ceperley_2016, place={Cambridge}, title={Interacting Electrons: Theory and Computational Approaches}, publisher={Cambridge University Press}, author={Martin, Richard M. and Reining, Lucia and Ceperley, David M.}, year={2016}}

@article{PhysRevB.82.165111,
  title = {Adiabatic and nonadiabatic phonon dispersion in a Wannier function approach},
  author = {Calandra, Matteo and Profeta, Gianni and Mauri, Francesco},
  journal = {Phys. Rev. B},
  volume = {82},
  issue = {16},
  pages = {165111},
  numpages = {16},
  year = {2010},
  month = {Oct},
  publisher = {American Physical Society},
  doi = {10.1103/PhysRevB.82.165111},
  url = {https://link.aps.org/doi/10.1103/PhysRevB.82.165111}
}

@article{PhysRevB.111.024307,
  title = {Exact formula with two dynamically screened electron-phonon couplings for positive phonon-linewidths approximations},
  author = {Stefanucci, Gianluca and Perfetto, Enrico},
  journal = {Phys. Rev. B},
  volume = {111},
  issue = {2},
  pages = {024307},
  numpages = {5},
  year = {2025},
  month = {Jan},
  publisher = {American Physical Society},
  doi = {10.1103/PhysRevB.111.024307},
  url = {https://link.aps.org/doi/10.1103/PhysRevB.111.024307}
}

@article{PhysRevX.13.041009,
  title = {Phonon Self-Energy Corrections: To Screen, or Not to Screen},
  author = {Berges, Jan and Girotto, Nina and Wehling, Tim and Marzari, Nicola and Ponc\'e, Samuel},
  journal = {Phys. Rev. X},
  volume = {13},
  issue = {4},
  pages = {041009},
  numpages = {29},
  year = {2023},
  month = {Oct},
  publisher = {American Physical Society},
  doi = {10.1103/PhysRevX.13.041009},
  url = {https://link.aps.org/doi/10.1103/PhysRevX.13.041009}
}

@article{PhysRevLett.68.3603,
  title = {Dielectric tensor, effective charges, and phonons in \ensuremath{\alpha}-quartz by variational density-functional perturbation theory},
  author = {Gonze, Xavier and Allan, Douglas C. and Teter, Michael P.},
  journal = {Phys. Rev. Lett.},
  volume = {68},
  issue = {24},
  pages = {3603--3606},
  numpages = {0},
  year = {1992},
  month = {Jun},
  publisher = {American Physical Society},
  doi = {10.1103/PhysRevLett.68.3603},
  url = {https://link.aps.org/doi/10.1103/PhysRevLett.68.3603}
}

@article{PhysRevLett.101.096802,
  title = {Temperature-Dependent Transport in Suspended Graphene},
  author = {Bolotin, K. I. and Sikes, K. J. and Hone, J. and Stormer, H. L. and Kim, P.},
  journal = {Phys. Rev. Lett.},
  volume = {101},
  issue = {9},
  pages = {096802},
  numpages = {4},
  year = {2008},
  month = {Aug},
  publisher = {American Physical Society},
  doi = {10.1103/PhysRevLett.101.096802},
  url = {https://link.aps.org/doi/10.1103/PhysRevLett.101.096802}
}

@article{PhysRevLett.100.016602,
  title = {Giant Intrinsic Carrier Mobilities in Graphene and Its Bilayer},
  author = {Morozov, S. V. and Novoselov, K. S. and Katsnelson, M. I. and Schedin, F. and Elias, D. C. and Jaszczak, J. A. and Geim, A. K.},
  journal = {Phys. Rev. Lett.},
  volume = {100},
  issue = {1},
  pages = {016602},
  numpages = {4},
  year = {2008},
  month = {Jan},
  publisher = {American Physical Society},
  doi = {10.1103/PhysRevLett.100.016602},
  url = {https://link.aps.org/doi/10.1103/PhysRevLett.100.016602}
}

@article{PhysRevB.77.041409,
  title = {Interplay of Coulomb and electron-phonon interactions in graphene},
  author = {Basko, D. M. and Aleiner, I. L.},
  journal = {Phys. Rev. B},
  volume = {77},
  issue = {4},
  pages = {041409},
  numpages = {4},
  year = {2008},
  month = {Jan},
  publisher = {American Physical Society},
  doi = {10.1103/PhysRevB.77.041409},
  url = {https://link.aps.org/doi/10.1103/PhysRevB.77.041409}
}

@article{PhysRevLett.92.075501,
  title = {Phonon Dispersion in Graphite},
  author = {Maultzsch, J. and Reich, S. and Thomsen, C. and Requardt, H. and Ordej\'on, P.},
  journal = {Phys. Rev. Lett.},
  volume = {92},
  issue = {7},
  pages = {075501},
  numpages = {4},
  year = {2004},
  month = {Feb},
  publisher = {American Physical Society},
  doi = {10.1103/PhysRevLett.92.075501},
  url = {https://link.aps.org/doi/10.1103/PhysRevLett.92.075501}
}

@article{PhysRevB.76.035439,
  title = {Phonon dispersion of graphite by inelastic x-ray scattering},
  author = {Mohr, M. and Maultzsch, J. and Dobard\ifmmode \check{z}\else \v{z}\fi{}i\ifmmode \acute{c}\else \'{c}\fi{}, E. and Reich, S. and Milo\ifmmode \check{s}\else \v{s}\fi{}evi\ifmmode \acute{c}\else \'{c}\fi{}, I. and Damnjanovi\ifmmode \acute{c}\else \'{c}\fi{}, M. and Bosak, A. and Krisch, M. and Thomsen, C.},
  journal = {Phys. Rev. B},
  volume = {76},
  issue = {3},
  pages = {035439},
  numpages = {7},
  year = {2007},
  month = {Jul},
  publisher = {American Physical Society},
  doi = {10.1103/PhysRevB.76.035439},
  url = {https://link.aps.org/doi/10.1103/PhysRevB.76.035439}
}

@article{PhysRevB.80.085423,
  title = {Phonon surface mapping of graphite: Disentangling quasi-degenerate phonon dispersions},
  author = {Gr\"uneis, A. and Serrano, J. and Bosak, A. and Lazzeri, M. and Molodtsov, S. L. and Wirtz, L. and Attaccalite, C. and Krisch, M. and Rubio, A. and Mauri, F. and Pichler, T.},
  journal = {Phys. Rev. B},
  volume = {80},
  issue = {8},
  pages = {085423},
  numpages = {5},
  year = {2009},
  month = {Aug},
  publisher = {American Physical Society},
  doi = {10.1103/PhysRevB.80.085423},
  url = {https://link.aps.org/doi/10.1103/PhysRevB.80.085423}
}

@article{PhysRevLett.93.185503,
  title = {Kohn Anomalies and Electron-Phonon Interactions in Graphite},
  author = {Piscanec, S. and Lazzeri, M. and Mauri, Francesco and Ferrari, A. C. and Robertson, J.},
  journal = {Phys. Rev. Lett.},
  volume = {93},
  issue = {18},
  pages = {185503},
  numpages = {4},
  year = {2004},
  month = {Oct},
  publisher = {American Physical Society},
  doi = {10.1103/PhysRevLett.93.185503},
  url = {https://link.aps.org/doi/10.1103/PhysRevLett.93.185503}
}

@article{PhysRevX.15.021043,
  title = {Symmetry-Dependent Dielectric Screening of Optical Phonons in Monolayer Graphene},
  author = {Moczko, Lo\"{\i}c and Reichardt, Sven and Singh, Aditya and Zhang, Xin and Jouaiti, Elise and L\'opez, Luis E. Parra and Wolff, Joanna L. P. and Moghe, Aditi Raman and Lorchat, Etienne and Singh, Rajendra and Watanabe, Kenji and Taniguchi, Takashi and Majjad, Hicham and Romeo, Michelangelo and Gloppe, Arnaud and Wirtz, Ludger and Berciaud, St\'ephane},
  journal = {Phys. Rev. X},
  volume = {15},
  issue = {2},
  pages = {021043},
  numpages = {18},
  year = {2025},
  month = {May},
  publisher = {American Physical Society},
  doi = {10.1103/PhysRevX.15.021043},
  url = {https://link.aps.org/doi/10.1103/PhysRevX.15.021043}
}

@article{PhysRevB.84.115442,
  title = {Disorder-induced temperature-dependent transport in graphene: Puddles, impurities, activation, and diffusion},
  author = {Li, Qiuzi and Hwang, E. H. and Das Sarma, S.},
  journal = {Phys. Rev. B},
  volume = {84},
  issue = {11},
  pages = {115442},
  numpages = {16},
  year = {2011},
  month = {Sep},
  publisher = {American Physical Society},
  doi = {10.1103/PhysRevB.84.115442},
  url = {https://link.aps.org/doi/10.1103/PhysRevB.84.115442}
}

@Article{jsan2010085,
AUTHOR = {Paun, Maria-Alexandra and Sallese, Jean-Michel and Kayal, Maher},
TITLE = {Hall Effect Sensors Design, Integration and Behavior Analysis},
JOURNAL = {Journal of Sensor and Actuator Networks},
VOLUME = {2},
YEAR = {2013},
NUMBER = {1},
PAGES = {85--97},
URL = {https://www.mdpi.com/2224-2708/2/1/85},
ISSN = {2224-2708},
DOI = {10.3390/jsan2010085}
}

@misc{accpaper,
  title = {Coupled Dynamical Boltzmann Transport Equations with Long-range Electron-phonon and Electron-electron Interactions in 2D Materials},
  author = {Macheda, Francesco and Sohier, Thibault},
  year = {2026},
}

@misc{SM,
  title = {See Supplemental Material [url] for complementary analysis.}
}

@article{Perdew_1996,
  title = {Generalized Gradient Approximation Made Simple},
  author = {Perdew, John P. and Burke, Kieron and Ernzerhof, Matthias},
  journal = {Phys. Rev. Lett.},
  volume = {77},
  issue = {18},
  pages = {3865--3868},
  numpages = {0},
  year = {1996},
  month = {Oct},
  publisher = {American Physical Society},
  doi = {10.1103/PhysRevLett.77.3865},
  url = {https://link.aps.org/doi/10.1103/PhysRevLett.77.3865}
}

@article{Binci_2021,
  title = {First-principles theory of infrared vibrational spectroscopy of metals and semimetals: Application to graphite},
  author = {Binci, Luca and Barone, Paolo and Mauri, Francesco},
  journal = {Phys. Rev. B},
  volume = {103},
  issue = {13},
  pages = {134304},
  numpages = {10},
  year = {2021},
  month = {Apr},
  publisher = {American Physical Society},
  doi = {10.1103/PhysRevB.103.134304},
  url = {https://link.aps.org/doi/10.1103/PhysRevB.103.134304}
}

@article{Laturia_2018,
	author = {Laturia, Akash and Van de Put, Maarten L. and Vandenberghe, William G.},
	date = {2018/03/08},
	doi = {10.1038/s41699-018-0050-x},
	id = {Laturia2018},
	isbn = {2397-7132},
	journal = {npj 2D Materials and Applications},
	number = {1},
	pages = {6},
	title = {Dielectric properties of hexagonal boron nitride and transition metal dichalcogenides: from monolayer to bulk},
	url = {https://doi.org/10.1038/s41699-018-0050-x},
	volume = {2},
	year = {2018}}

@Article{Neumann2015,
author={Neumann, C.
and Reichardt, S.
and Venezuela, P.
and Dr{\"o}geler, M.
and Banszerus, L.
and Schmitz, M.
and Watanabe, K.
and Taniguchi, T.
and Mauri, F.
and Beschoten, B.
and Rotkin, S. V.
and Stampfer, C.},
title={Raman spectroscopy as probe of nanometre-scale strain variations in graphene},
journal={Nature Communications},
year={2015},
month={Sep},
day={29},
volume={6},
number={1},
pages={8429},
issn={2041-1723},
doi={10.1038/ncomms9429},
url={https://doi.org/10.1038/ncomms9429}
}
\end{document}

\title{Supplementary Material for `Electronic transport in BN-encasulated graphene limited by remote phonon scattering'}

\author{K. Dinar}
    \affiliation{\lcc}
    \affiliation{Aix Marseille Univ, CNRS, CINaM, AMUTech, Marseille, France.}

\author{F. Macheda}
\affiliation{Dipartimento di Scienze e Metodi dell’Ingegneria,
University of Modena and Reggio Emilia, Reggio Emilia, Italy}
\affiliation{Dipartimento di Fisica, Sapienza Università di Roma, Roma, Italy}
\author{A. Guandalini}
\affiliation{Dipartimento di Fisica, Sapienza Università di Roma, Roma, Italy}
\author{M. Paillet}
\affiliation{\lcc}
\author{C. Consejo}
\affiliation{\lcc}
%\author{G. Sigu}
%\affiliation{\lcc}
\author{F. Teppe}
\affiliation{\lcc}
\author{B. Jouault}
\affiliation{\lcc}
\author{T. Sohier}
\affiliation{\lcc}
\author{S. Nanot}
\affiliation{\lcc}

\maketitle

This supplementary material is divided in the following sections:
in Sec. \ref{sec:expdet} we present the experimental details that are not contained in the main text of this article (device image, mobility and reproducibility); in Sec. \ref{sec:phfreq} we present a detailed calculation of the graphene $A'_1$ phonon frequency; in Sec. \ref{sec:calc} we present the computation of graphene resistivity obtained by changing the temperature, the number of BN encapsulating layers and the strength of the anharmonic coupling; finally, in Sec. \ref{app:fitProc} we discuss in detail the fitting procedure used in this work  to extract the proper values of the phonon frequency and electron-phonon coupling strength, and in Sec. \ref{sec:additivity}, why a Boltzmann approach to fit the phonon contribution does not lead to proper values.

\section{Experimental details}
\label{sec:expdet}

\begin{figure}
    \centering
    \includegraphics[width=0.98\linewidth]{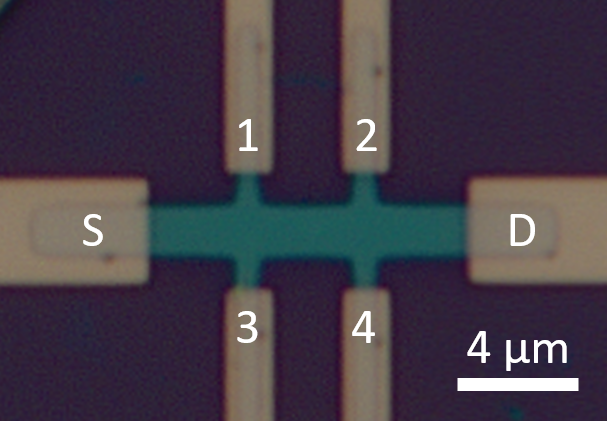}
    \caption{Optical image of the Hall bar device presented in the main text.}
    \label{fig:Opt_image}
\end{figure}

The device fabrication is detailed in the main text and its appendix. We show an optical image taken right before the measurements in Fig. \ref{fig:Opt_image}.

\begin{figure}
    \centering
    \includegraphics[width=1\linewidth]{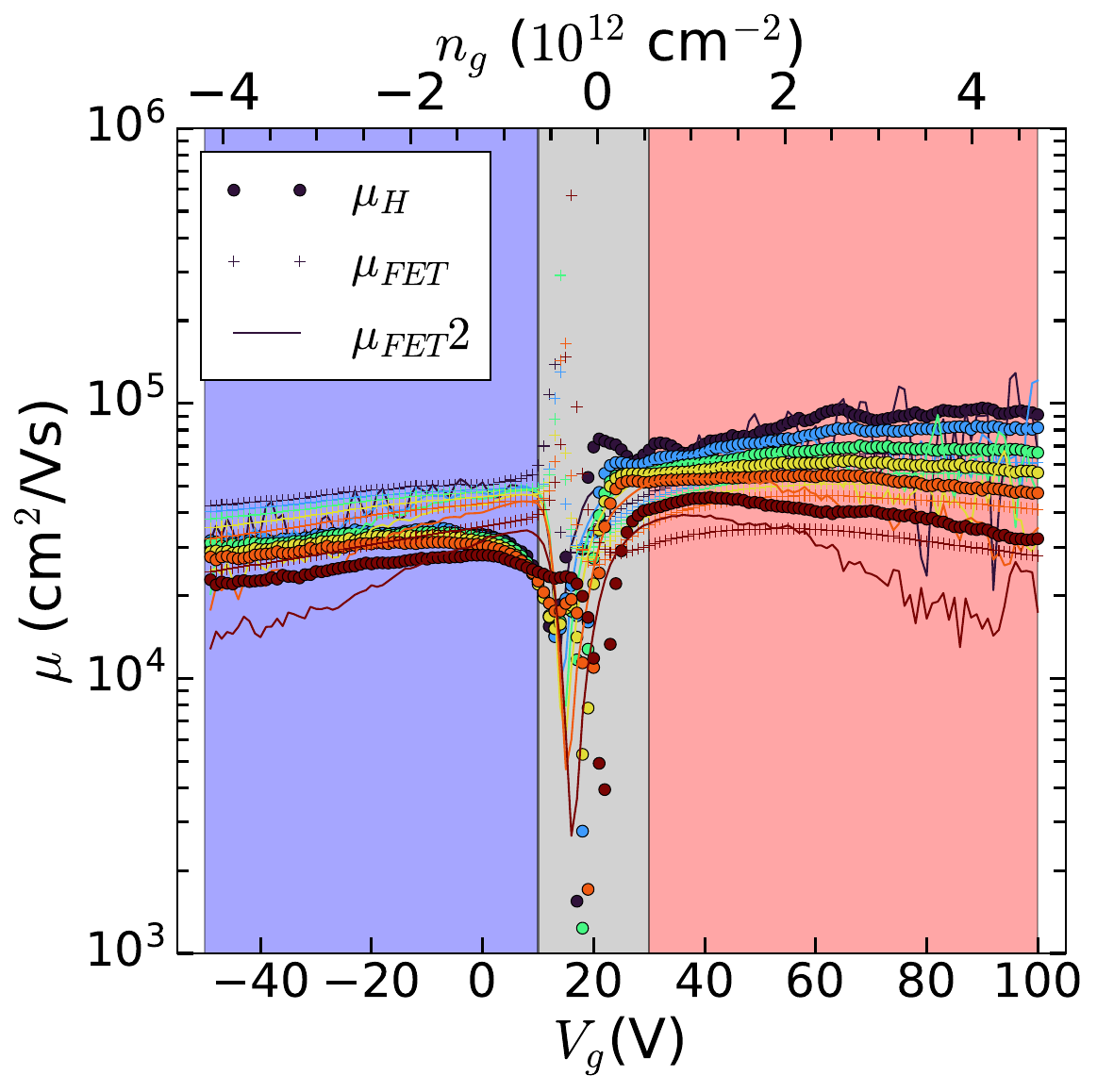}
    \caption{Comparison of Hall mobility (points) and FET mobility (crosses and lines) determined using two methods (see text) from 10 K (blue curves) to 300 K (wine red curves). The gray area is where $n_{\rm{H}} \neq n_{\rm{g}}$, while the blue and red ones are for $\varepsilon_{\rm{F}} \ge 100\, \text{meV}$. Colors are the same as in Fig. 4 in the main text. }
    \label{fig:Mobility_exp}
\end{figure}

In order to confirm that the carrier density is properly extracted and that transport is dominated by one-carrier diffusive transport, we compared the Hall and field-effect mobilities obtained using $\mu_{\rm{H}} = \sigma/n_{\rm{H}}e$, $\mu_{\rm{FET}} = \sigma/n_{\rm{g}}e$ and $\mu_{{\rm{FET}},2} = \frac{1}{Ce}d\sigma/dV_{\rm{g}}$. One can see in Fig. \ref{fig:Mobility_exp} that the 3 methods give the same order of magnitude for doping away from the CNP. While the last method is noisier due to a choice of limited smoothing and averaging (to avoid other analysis artifacts), $\mu_{\rm{H}}$ and $\mu_{\rm{FET}}$ give the same gate-independent behavior. A slight difference can be observed between the two methods. This is probably due to the slight non-linearity of $n_{\rm{H}}(V_{\rm{g}})$ which can be explained by the device inhomogeneities and other corrections due to the geometry of the Hall bar and carrier scattering \cite{jsan2010085,Macheda2020}. We decided to report the most conservative values in the main text (\emph{e.g.} $\mu_{\rm{holes}}\simeq 20{,}000\,\text{cm}^2\text{/Vs}$ at $T = 300\,\text{K}$).

\begin{figure*}
    \centering
    \includegraphics[width=\linewidth]{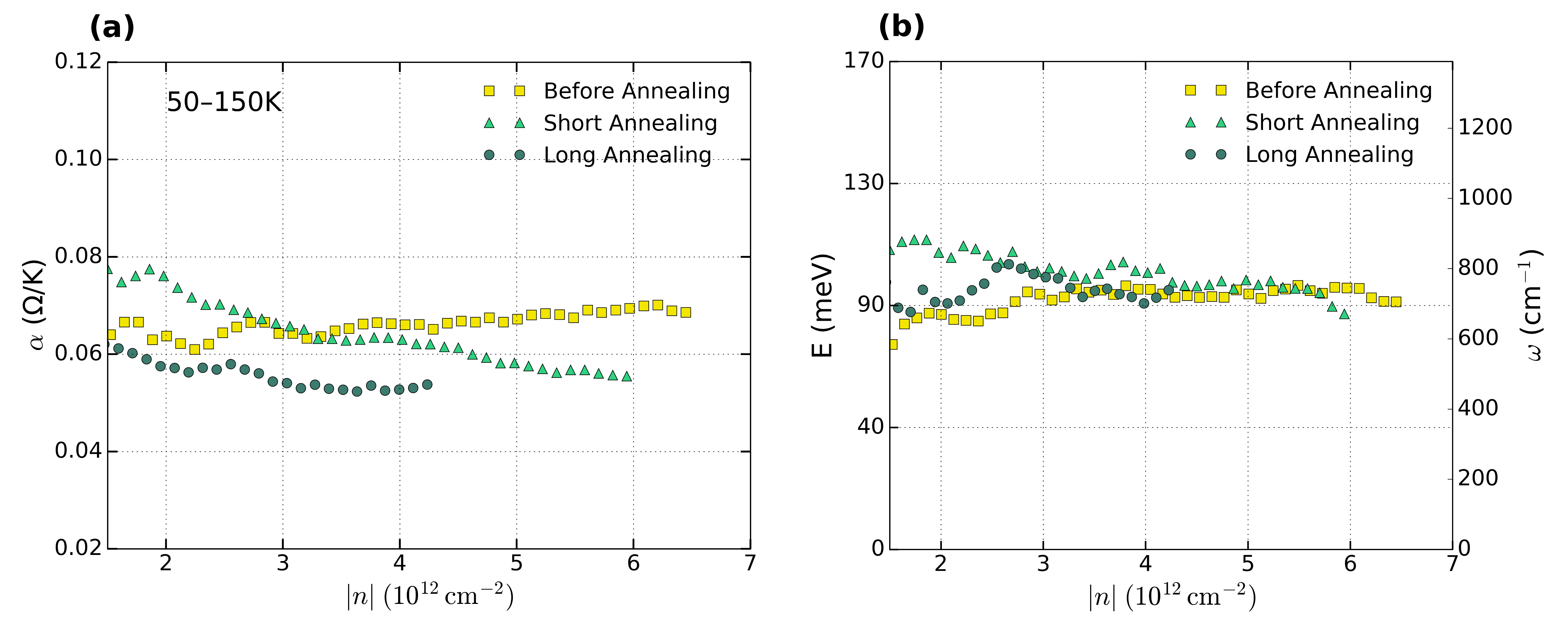}
    \caption{(a) Slope $\alpha = d\rho/dT$, extracted between 50-150\,K, plotted versus carrier density for three thermal cycles: before annealing (yellow squares), after short annealing (green triangles), and after long annealing (dark green circles). These data correspond to holes; electrons show the same overall trend. (b) Optical phonon energy extracted using \ref{eq:rhowrong}, demonstrating a consistent energy scale across different measurements.}
    \label{fig:Reproducibility}
\end{figure*}

The measurements shown in this work have been reproduced by measuring this sample before an annealing in vacuum at $600$ K and after two annealings (of 2 and 10 minutes respectively). The data shown in the main text correspond to the last annealing. After each annealing, the residual doping was reduced, with $V_{\rm{CNP}}$ tending toward 0 V, and similarly the residual carrier density $n_0$ was also reduced. This showed that the device was more intrinsic and less inhomogeneous. Nevertheless, the phonon contribution to the resistivity is still similar for high enough carrier densities, as illustrated in Fig. \ref{fig:Reproducibility}.b. We obtain the same energy-indepedent coupling strength $\alpha$ between electrons and acoustic phonons, as well as similar optical phonon energies. Note that the fitting procedure used in Fig. \ref{fig:Reproducibility}.b gives an inexact value for the phonon energy. This is due to an inaccurate approach detailed in section \ref{sec:additivity} and does not affect the conclusions about reproducibility.

\section{Graphene phonon frequency at zone-border with doping and BN-encapsulation}
\label{sec:phfreq}
The intrinsic transport properties of graphene are mainly determined by scattering with acoustic phonons, and with the $E_{2g}$ optical phonons at $q$ around $\Gamma$ and the $A'_1$ phonon at $q$ around $K$. The frequency of the $E_{2g}$ phonons at $q=\Gamma$ are only marginally sensitive to doping and encapsulation, while at $q=K$ the phonon dispersion strongly depends on environmental conditions \cite{Guandalini2025}. In this section, we discuss the value of the optical phonon frequency in graphene at $q=K$ under different experimental conditions.

\textit{Theory---}We briefly summarize the theoretical framework developed in Refs. \cite{Guandalini2025,Guandalini2025a,Caldarelli2025}. In the harmonic approximation and assuming both that the electronic self-energy is real and that Migdal vertex corrections are small, the phonon self-energy can be written as an electronic response function. In our theoretical framework, we indeed employ a real electronic self-energy to determine single-particle self-energies via the screened-exchange (SX) approximation, i.e. we consider $\Sigma(W)$ where $W$ is the frequency independent screened Coulomb interaction obtained in the random-phase-approximation (RPA) \cite{Martin_Reining_Ceperley_2016}. Then, we choose the same form of $W$ as the direct-kernel of the Bethe-Salpeter equation (BSE), i.e. the equation that determines any electronic response function when contracted with appropriate vertexes. Within this framework, the phonon self-energy is obtained by solving the BSE, and then contracting the four-points dressed electron-hole propagators pairwise with the electron-phonon vertexes. Thanks to the partially screened-partially screened rewriting of the phonon self-energy \cite{PhysRevB.82.165111,Caldarelli2025}, generalizing the screen-screen approach \cite{PhysRevB.111.024307,PhysRevX.13.041009,PhysRevLett.68.3603}, we can isolate the many-body correction (beyond DFT) to the phonon self-energy, which we call $\Pi^{\rm BSE}$ here, and write it in terms of single-particle SX energies, DFT electron-phonon vertexes and $W$ (see Eq. 8, App. C and SM of Ref. \cite{Guandalini2025}). We then describe all the above ingredients with a tight binding model containing only $\pi/\pi^*$ graphene bands, with coefficients fitted over DFT results, to finally have a fast though accurate way to compute the phonon self-energy.

\textit{Phonon self-energy evaluation---}
We write the phonon self energy as
\begin{align}
\Pi_{s\alpha,r\beta}(q,\omega)\approx\Pi^{\rm DFT}_{s\alpha,r\beta}(q)-\Pi^{0,\rm DFT}_{s\alpha,r\beta}(q)+\Pi^{0,\rm BSE}_{s\alpha,r\beta}(q,\omega),
\end{align}
where
\begin{itemize}
\item $\Pi_{s\alpha,r\beta}(q)$ is our approximation to the phonon self-energy;
\item $\Pi^{\rm DFT}_{s\alpha,r\beta}(q)$ is the frequency independent DFT phonon self-energy, coinciding with the DFT force constant matrix $C_{s\alpha,r\beta}(q)$;
\item $\Pi^{0,\rm DFT}_{s\alpha,r\beta}(q)$ is the frequency independent DFT contribution to the self-energy due to the $\pi/\pi^*$ bands, that include all the non-analytical Khon anomaly behaviours. In practice, we evaluate $\Pi^{0,\rm DFT}_{s\alpha,r\beta}(q,\omega)$ with our tight binding framework, using DFT single-particle energies and couplings;
\item $\Pi^{0,\rm BSE}_{s\alpha,r\beta}(q,\omega)$ is the many-body correction to the dynamical matrix due to the $\pi/\pi^*$ bands, as evaluated in out tight-binding framework solving the BSE equation, that adds back non-analytical behaviors near the Khon anomaly.
\end{itemize}
Practically, we first compute $\Pi^{\rm DFT}_{s\alpha,r\beta}(q_i)$ with Quantum Espresso (QE) \cite{Carnimeo2023} on a set of $\{q_i\}$ points chosen to study the phonon frequencies $\omega_{\mu q_i}$, and contextually determine the eigenvectors $e^{\mu}_{q_i}$, where $\mu$ is the phonon branch index. Then, we subtract $\Pi^{0,\rm DFT}_{s\alpha,r\beta}(q_i)$, carefully computed at the same electronic temperature as in the DFT calculation, and add back $\Pi^{0,\rm BSE}_{s\alpha,r\beta}(q_i,\omega)$ obtained via the solution of the BSE equation. We give the computations details for the solution of the BSE below. Importantly, $\Pi^{0,\rm BSE}_{s\alpha,r\beta}(q_i,\omega)$ should in principle be computed on a fine frequency grid, and one should identify the renormalized phonon frequency looking at the phonon spectral function \cite{Guandalini2025}
\begin{equation}
A(\omega)={-}\sum_{s\alpha}{\rm Im}\left[\frac{\omega}{\pi} G_{s\alpha,s\alpha}(\omega)\right].
\end{equation}
In practice, we compute  $\Pi^{0,\rm BSE}_{s\alpha,r\beta}(q_i,\omega)$ for $\omega=0$ (statical) and $\omega=\omega^{\rm DFT}_{A'_1}$ (dynamical), where $\omega^{\rm DFT}_{A'_1}$ is the zone border $A'_1$ phonon frequency as obtained with DFT calculations. The error of this procedure is small, as tested in Ref. \cite{Guandalini2025}.

After this step, we obtain our approximation of the electronic self-energy $\Pi_{s\alpha,r\beta}(q_i,\omega)$ (statical or dynamical), and the phonon frequency is determined diagonalizing its Hermitian part, and taking the square root.

\textit{Doping and encapsulation---} Doping and encapsulation enter in the determination of the phonon self-energy mostly via $W$. In particular, doping modifies the irreducible polarizability of graphene $\chi^0$, and in turn $W$ via the RPA approximation. Similarly, encapsulation is taken into account by considering the effect of a dielectric environment on $W$ via the semi-analytical model of Ref. \onlinecite{Trolle2017}. In our case, we just consider two semi-infinite dielectric planes surrounding graphene, which is modeled as a cavity of thickness 3.50$\AA$. Therefore, we introduce both doping and encapsulation effects only in the computation of $\Pi^{0,\rm BSE}_{s\alpha,r\beta}(q_i,\omega)$.

Notice that we could use the VED model to evaluate BN remote electronic screening, but it would be heavier computationally and harder to interface with the code developed for the phonon frequency computations; in this respect, the semi-analytical model is expected to describe screening in a satisfactory way, with less computational effort.

\begin{figure*}
    \centering
    \includegraphics[width=\linewidth]{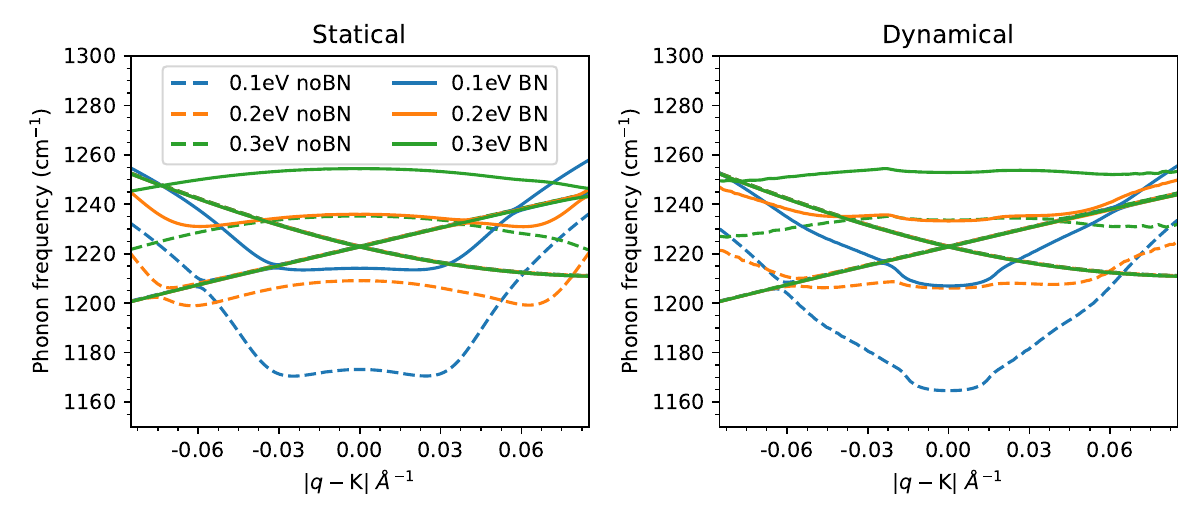}
    \caption{Phonon dispersion around the $q=K$ point of graphene, with statical (left panel) or dynamical (right panel) calculations at different dopings (blue, orange and green for 0.1, 0.2 and 0.3eV respectively) and in absence (dashed lines) or presence (continuous lines) of encapsulation. The lines that change with doping and BN represent the phonon that, exactly at K, has $A'_1$ character. The other two crossing bands are other branches which are not affected by the environment.}
    \label{fig:phdisp}
\end{figure*}

\textit{Computational details---}
The phonon dispersions shown in Fig.~\ref{fig:phdisp} have been obtained, with electron-electron (e-e) interaction effects included, as described in Ref.~\cite{Guandalini2025}.
The static DFT dynamical matrix $C^{\rm DFT}_{s\alpha,r\beta}$ have been obtained with the Quantum-ESPRESSO package~\citep{Carnimeo2023} with the PBE approximation~\cite{Perdew_1996}. We adopt norm-conserving pseudopotentials to model the electron-ion interaction and a kinetic energy cutoff of $90$ Ry.
The e-e interaction and/or dynamical corrections $\Delta C(z)$ are computed using the interacting, $\pi$-band, tight-binding approach of Ref.\ \onlinecite{Guandalini2025a}, where the screened interaction $\mathcal{W}$ is self-consistently evaluated with the band structure~\cite{Guandalini2025a}.
The parameters of the model are fitted on ab initio calculations.

The eigenvalue equation of the dynamical matrices are solved with the Sternheimer equation explained in Ref.~\onlinecite{Caldarelli2025}, with linear mixing and a convergence threshold of the force constants of $10^{-6}$.
We used telescopic k grids~\cite{Binci_2021} that in the dense region around K are equivalent to a $1448\times1448$ uniform grid, while in the coarse region they correspond to a $90\times90$ uniform grid. The parameters are $p=3$, $\mathcal{N}=25$, $l=6$ and $L=10$ in the DFT and tight-binding calculations.
The electronic temperature is set to $300$ K.
hBN encapsulation effects on $\mathcal{W}$ have been obtained via an anisotropic generalization of the three-layer dielectric model described in Ref.~\onlinecite{Trolle2017}, with $t =3.5$ \AA the graphene thickness and $\epsilon^{\parallel} = 4.98$ and $\epsilon^{\perp} = 3.03$ the static parallel and perpendicular dielectric constants of hBN, taken from Ref.~\onlinecite{Laturia_2018}. Dynamical effects are computed in the neighborhood of K evaluating the dynamical matrix at the static frequencies  $\omega_{\textrm{K}} = 1128$ cm$^{-1}$ of freestanding material. We verified slight changes in the frequency evaluation do not alter the final results.
Due to the finite k-point mesh, we replace the $0^+$ limit in the definition of the phonon Green's function with a finite damping $\eta = 10$ meV.

\textit{Results---} In Fig. \ref{fig:phdisp} we present the phonon dispersion of graphene around the $K$ point. In particular, on the horizontal axis we present the distance from $q$ to the $K$ point, in \AA. We present the results for statical calculations (left panel) and dynamical ones (right panel), at different dopings (blue, orange and green for 0.1, 0.2 and 0.3 eV respectively) and in absence (dashed lines) or presence (continuous lines) of encapsulation. We can immediately see that doping and encapsulation effects affect only the $A'_1$ phonon, as expected. The effect of doping is to reduce the strength of the Khon anomaly in a region of momentum space which is proportional to the Fermi wavevector, which is $k_{\rm F}=0.018,0.037,0.055$ \AA$^{-1}$ for the three levels of doping selected here. Further, encapsulation produces an almost rigid blue-shift of the phonon frequencies, in agreement with what found experimentally in Ref. \cite{Venanzi2023} when comparing suspended and encapsulated samples. Finally, dynamical effects only slightly modify the phonon dispersion around the $K$ point, due to the creation of electron-hole pairs. In conclusion, we find that the zone-border frequency of the $A'_1$ phonon in BN-encapsulated graphene is roughly position around $\sim 1225$ cm$^{-1}$, with contained changes in the doping regime considered. This value is also conservatively close to the experimental results  \cite{PhysRevLett.92.075501,PhysRevB.76.035439,PhysRevB.80.085423} and ab initio DFT calculations performed on graphite \cite{PhysRevLett.93.185503}.
\section{VED results}
We here report the results for the Van der Waals heterostructure method (VED), introduced in Ref. \cite{Macheda2024b}, and expanded in an accompanying paper \cite{accpaper} to address transport properties. We refer to the above references for an extended account of the computational methodology and computational details, that are only briefly described in the following.
\label{sec:calc}
\textit{Computation---} We remind that we use Matthiesen's rule to add intrinsic scattering mechanisms and remote ones. In addition to what reported in Ref. \onlinecite{accpaper}, we choose the form of the anharmonic scattering rate of BN phonons to be the constant \cite{accpaper}
\begin{align}
\tau^{-1}_{\rm anh}=\tau^{-1}_{\rm anh, BN}= \tau^{-1}_{\rm LO, BN}+ \tau^{-1}_{\rm ZO, BN}
\label{eq:taum1phen}
\end{align}
with $\tau^{-1}_{\rm ZO, BN}=\tau^{-1}_{\rm LO, BN}/10$ as found from TDEP calculations \cite{Hellman2013}, where `anh' stands for `anharmonic'. Our choice is consistent with having an anharmonic scattering rates which integrates to the integrated anharmonic interaction of the system \cite{accpaper}:
\begin{align}
\int d\omega \tau^{-1}_{\rm ANH}(\mathbf{q},\omega)= \tau^{-1}_{\rm anh} \int d\omega  \mathcal{F}(\mathbf{q},\omega) = \nonumber\\
\rm{NBN} \times \left[ \tau^{-1}_{\rm LO, BN}+ \tau^{-1}_{\rm ZO, BN}  \right],
\label{eq:taum1tot}
\end{align}
where NBN is the number of BN layers on each side of graphene---this is the same number of IR active modes \cite{Macheda2024b}. Notice that our definition for the phonon content $\mathcal{F}$, described in Ref. \onlinecite{accpaper}, is not additive with respect to phonons excitation. In other words, if we integrate $\mathcal{F}$ only around the ZO frequency, we do not get NBN/2. Consequently, the attribution of a frequency dependent $\tau^{-1}_{\rm ANH}(\omega)$ that associates the correct anharmonic decay to each phonon is not possible, and the only constraint that we can satisfy is the sum rule over the total anharmonic scattering rate. We leave the solution of this problem for future investigations. Notice also that Eq. \eqref{eq:taum1tot} does not mean that the total anharmonic scattering rate of the bulk is infinite: for $\rm{NBN} \rightarrow \infty$, every discrete mode is mapped as a point of the $q_z$ dispersion of the bulk phonon, and our approximation amounts to say that the anharmonic scattering rate does not depend on $q_z$; the density of anharmonic scattering per unit state is then constant.

Below, we will also compare with the choice
\begin{align}
 \tau^{-1}_{\rm ANH}(\mathbf{q},\omega)=
 \begin{cases}
 \infty \quad \textrm{if} \, \mathcal{F}(\mathbf{q},\omega) \neq0\\
 0 \quad \textrm{if} \, \mathcal{F}(\mathbf{q},\omega) =0
\end{cases}.
\label{eq:taum1inf}
\end{align}

\begin{figure}
    \centering
    \includegraphics[width=\linewidth]{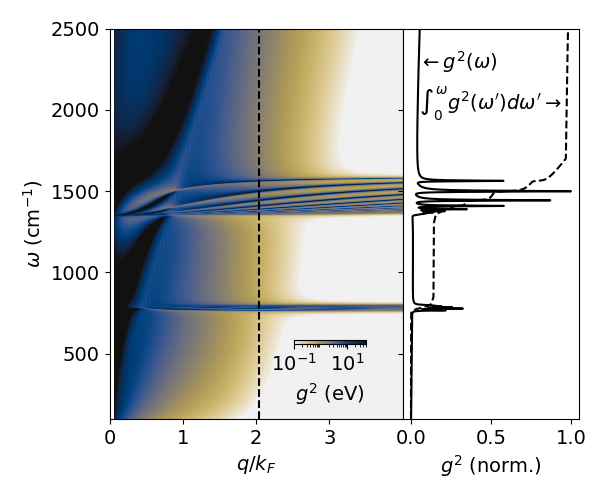}
    \caption{Electron-electrodynamic mode coupling in graphene encapsulated by 20 layers of BN on each side. The left panel show $g^2(q,\omega)$ in logarithmic colorscale. The right panel shows $g^2(\omega)$ at fixed $q \sim 2 k_F$ corresponding to the dashed in the left panel, as well as the cumulative integral $\int_0^{\omega} g^2(\omega')d\omega'$, both normalized by their highest value.}
    \label{fig:gw}
\end{figure}

\textit{Coupling strength between electrons LO and ZO---} The coupling between electrons and the LO and ZO modes is depicted in Fig. \ref{fig:gw}, following what discussed in Refs. \cite{Macheda2024b,accpaper}. On the left panel, we represent the dispersions of all the electrodynamical modes for a 20BN-Gr-20BN heterostructure in the full momentum-frequency plane. On the right panel, we plot the coupling $g(q,\omega)$, for the momentum corresponding to the vertical dashed line presented in the left panel---we call it $g(\omega)$ in the figure for brevity. We also represents its cumulative integral with a dashed line. As evident, LO modes present strong interaction peaks, higher than ZO ones. Nonetheless, ZO modes are situated at lower frequencies and end up contributing much importantly to the low temperature resistivity of graphene.

\textit{Impurity scatterers effects on intrinsic resistivity --- }
Impurities are responsible for the limit of the resistivity towards low temperatures $\rho(T=0) = \rho_0$. In our transport simulations, it is accounted for by defining an energy-independant scattering time $\tau_0$ that, taken alone, would yield $\rho_0$. The phonon and impurity contributions are then added following Mathiessen's rule, that is by summing the inverse of scattering times. Because this procedure is not equivalent to adding the corresponding resistivities, the values of $\rho(T)-\rho_0$ as a function of temperature will slightly depend on the impurities. To account for this in our simulation, we compute $\rho(T)-\rho_0$ for a range of $\rho_0$ corresponding to what is observed experimentally. This corresponds to the grey shading in Fig. 2 of the main text, and we can clearly see that the associated variation of resistivity cannot explain the deviation from the model including only intrinsic phonons.

\begin{figure*}
    \centering
    \includegraphics[width=0.65\linewidth]{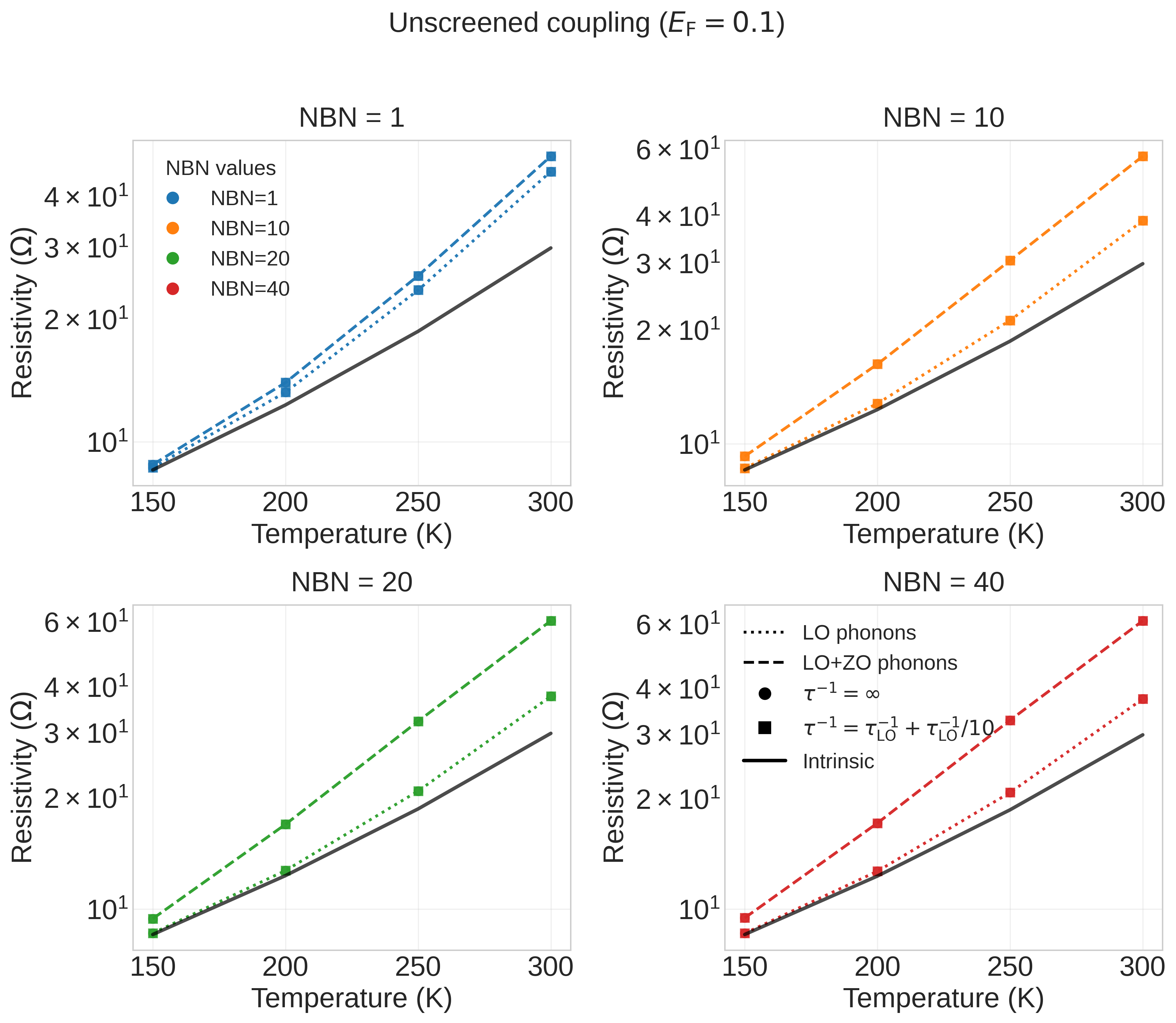}
    \caption{Graphene resistivity in presence of an unscreened remote phonon coupling (i.e. disregarding the effect of graphene free carriers on screening) as a function of the number of BN layers (NBN); we evaluate the effect of the remote coupling with BN ZO phonons by computing the resistivity due to remote BN LO phonon alone, and the one including both LO and ZO remote phonons. The data in this figure refer to $E_{\rm F}=0.1$eV.}
    \label{fig:bare0.1}
\end{figure*}

\begin{figure*}
    \centering
    \includegraphics[width=0.65\linewidth]{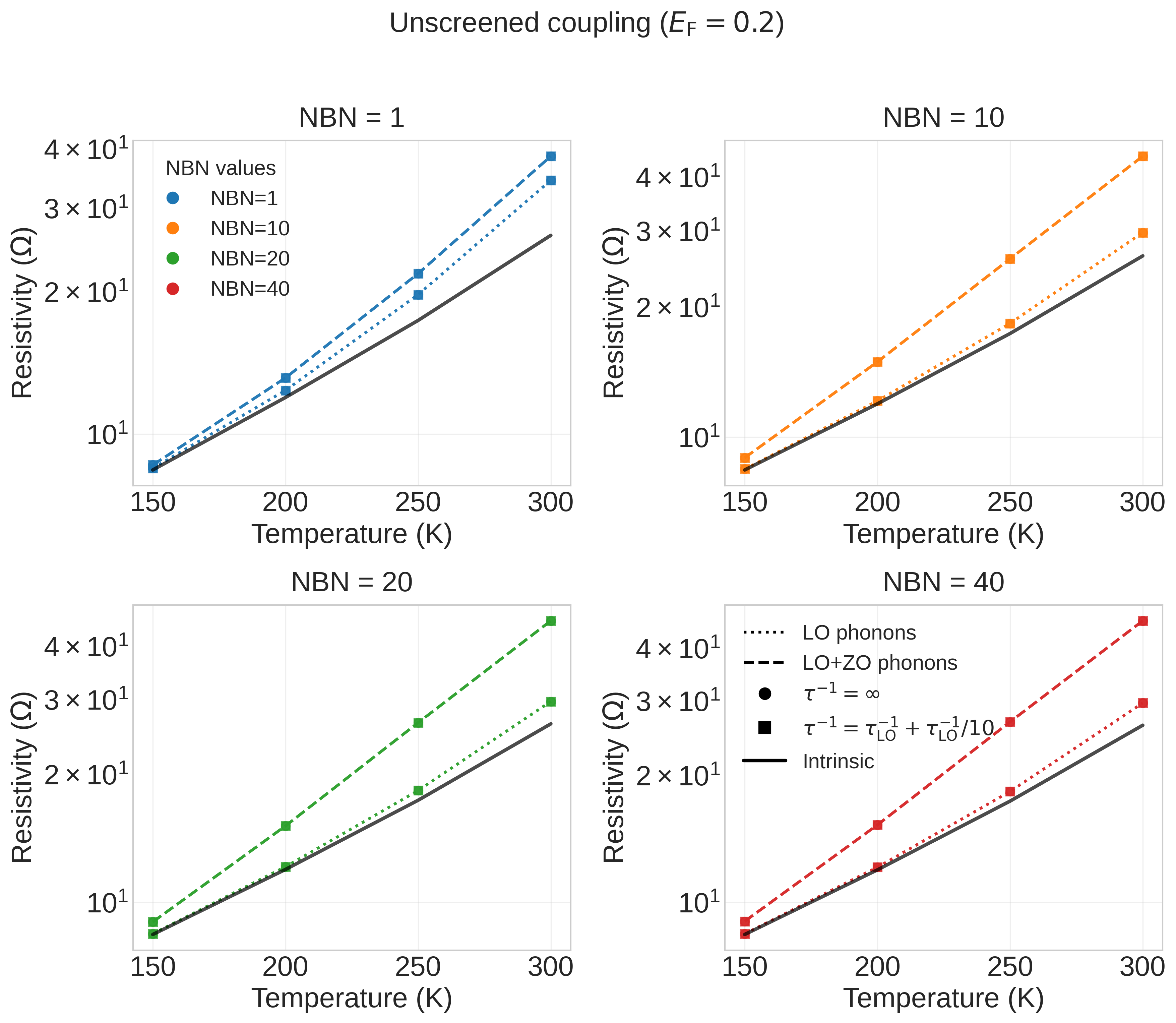}
    \caption{Same as Fig. \ref{fig:bare0.1} but for $E_{\rm F}=0.2$eV}
    \label{fig:bare0.2}
\end{figure*}

\begin{figure*}
    \centering
    \includegraphics[width=0.65\linewidth]{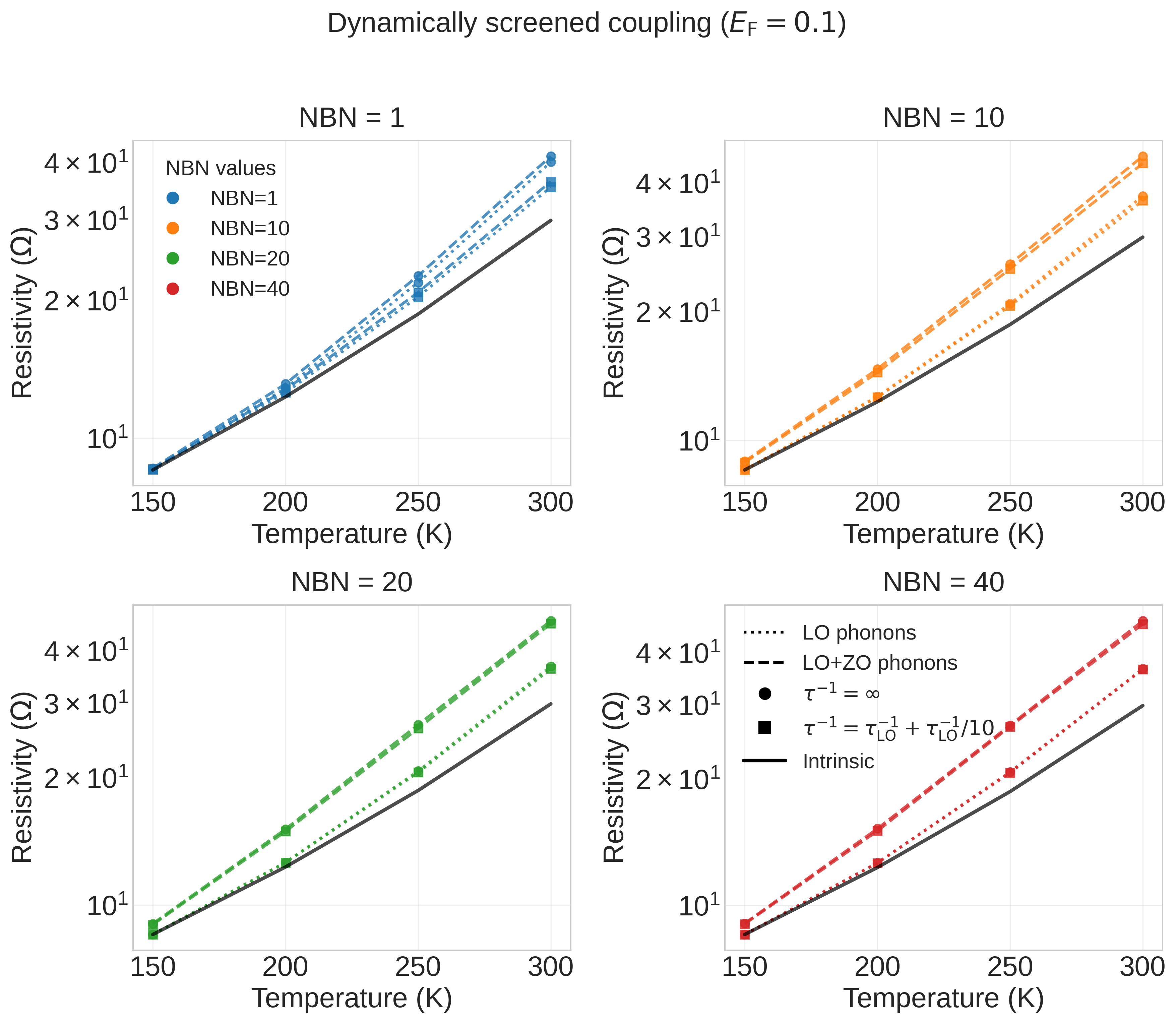}
    \caption{Same as Fig. \ref{fig:bare0.1} but for dynamically screened coupling (i.e. considering the effect of graphene free carriers on the screening of remote phonons). We further plot the resistivity value computed considering the anharmonic scattering via Eq. \ref{eq:taum1phen} and \ref{eq:taum1inf}. The data in this figure refer to $E_{\rm F}=0.1$eV.}
    \label{fig:dyn0.1}
\end{figure*}

\begin{figure*}
    \centering
    \includegraphics[width=0.65\linewidth]{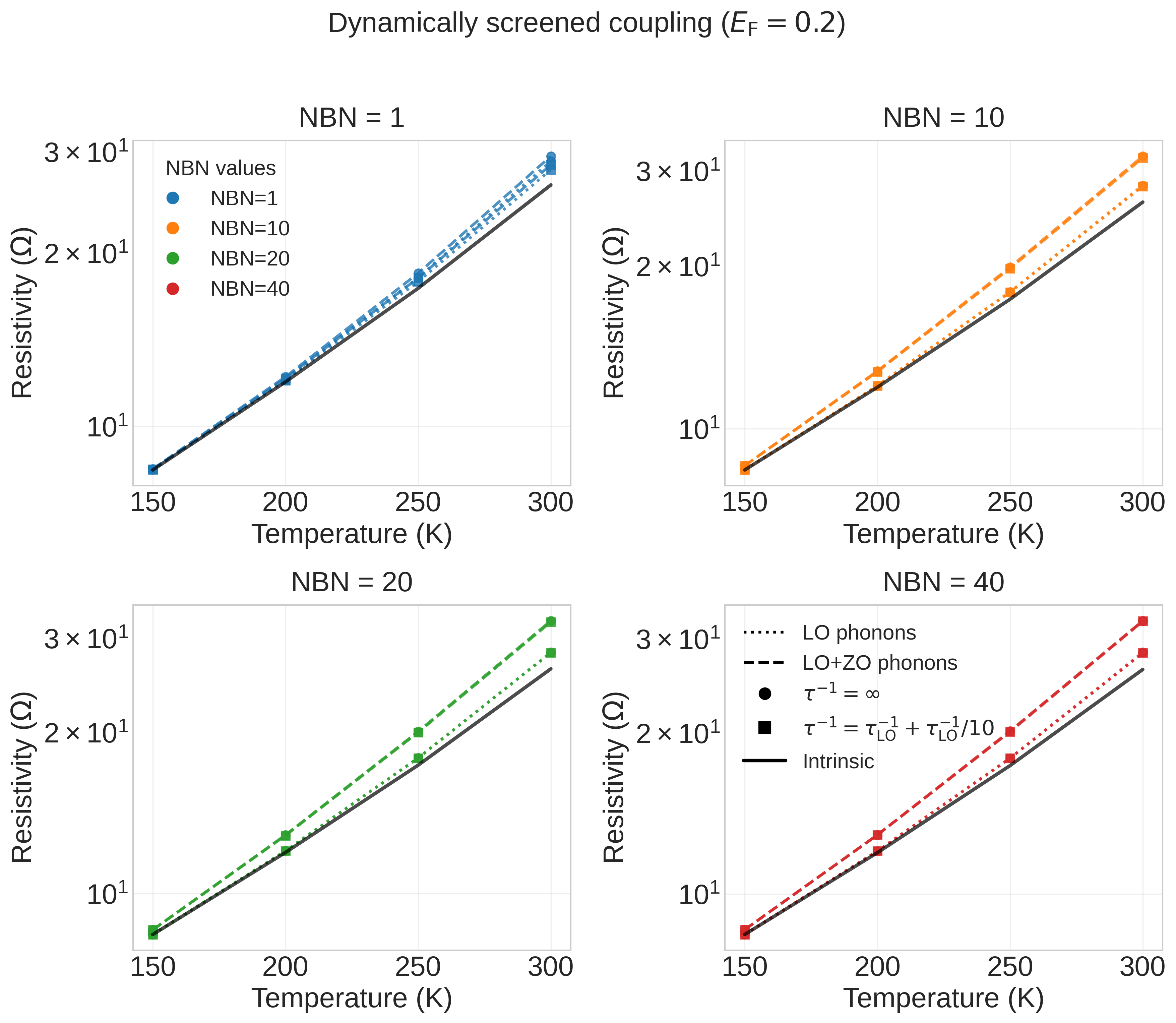}
    \caption{Same as Fig. \ref{fig:dyn0.1} but for $E_{\rm F}=0.2$}
    \label{fig:dyn0.2}
\end{figure*}

\textit{Influence of different configurational setups on graphene resistivity---} We study the influence of the Fermi level $E_{\rm F}$ and numbers of BN layers (NBN) on graphene resistivity. We first consider
the remote phonon coupling as unscreened by graphene electrons, i.e. we set it to the value it would have in a BN heterostructure composed of NBN layers without graphene, and solve the electronic Boltzman equation applying the Bloch's ansatz. This is meant to show the impact of the Fermi surface change and of the number of BN layers on resistivity, independently of screening effects. In particular, an increasing $E_{\rm F}$ allows both an increase of carriers and an enlargement of the phase space for optical phonon scattering, while an increase in NBN entails the transition of the remote phonon coupling from the two dimensional to the three dimensional value. The results are shown in Figs. \ref{fig:bare0.1} and \ref{fig:bare0.2}. Each figure refers to a given value of the Fermi energy (0.1,0.2), and presents four panel corresponding to different NBN (1,10,20,40). We further distinguish between calculations performed including only LO remote phonons (dotted lines) or both LO and ZO phonons (dashed lines); the instrinsic resistivity is plotted as a continuous line as reference. From the data, we draw three conclusions:
\begin{enumerate}
    \item The resistivity due to LO phonons alone lowers as NBN increases, while the resistivity due to LO+ZO phonons roughly remains constant;
    \item The importance of ZO phonons greatly increase with the increase of BN layers, saturating at around NBN=30. This is due to a coherent buildup of the electrostatic potential on the graphene layer coming from the remote ZO phonons;
    \item Resistivity decreases with increasing Fermi level, meaning that the growth in the number of carriers outweights the phase space increase for optical phonon scattering.
\end{enumerate}
All the above conclusions remain qualitatively valid when screening effects are now included in the calculations, as shown in Figs. \ref{fig:dyn0.1} and \ref{fig:dyn0.2}. In this case, though, the quantitative behavior of the resistivity as a function of Fermi level is further influenced by the free-electron scattering, which strongly increases alongside with the free-carriers density. Indeed, the remote coupling contribution to the total resistivity is far reduced with respect to the case where the remote interaction is taken as unscreened.

For the dynamically screened coupling case, we further consider the two different treatments for the anharmonic scattering of Eqs. \eqref{eq:taum1phen} (square symbols) and \eqref{eq:taum1inf} (circle symbols). We notice here that the difference between these two approaches is quite marked for the low NBN case, while it saturates to a negligible values when NBN increases. The reason is that the phonon content becomes a series of narrower peaks when NBN increases, and the peak height of the excitations thus does not have to decrease as 1/NBN to compensate normalization scaling; as a consequence, anharmonic scattering is more effective around the phonon excitation. We leave the discussion of this feature for future investigations. Given the small difference of the two methods for large NBN, in the main text we use the prescription of Eq. \ref{eq:taum1inf}.


\section{Fitting optical phonon contributions in experiments and theory}
\label{app:fitProc}
The scope of this section is to describe the fitting procedure used to extract BN's ZO optical phonon coupling to electrons. The general idea is to first isolate the contribution to the resistivity coming from all optical phonons, and then fit their coupling constants with electrons. Even though in principle we know the theoretical couplings, we perform a fitting procedure on both experimental and theoretical resistivities to compare exactly the same quantities.

To understand how to extract a reasonable optical phonon scattering strength, we first need to understand how couplings roughly reflect on resistivities. We therefore consider the electronic Boltzmann transport equation for graphene, solved within the Dirac cone approximation, following Ref. \cite{Sohier2014}. In particular, Appendix A of  Ref. \cite{Sohier2014} shows how the BTE can be solved by inverting a matrix indexed on energy. While that is easily done with minimal computing power, it can take more to compute the resistivity at several temperatures. In this work, we need a faster solution to inject into a curve fitting algorithm and fit the coupling to the optical phonons. The matrix inversion problem becomes trivial and very fast only if the corresponding matrix is diagonal.

%Treatment of acoustic phonons
As discussed in Appendix A of Ref. \cite{Sohier2014},  this is already the case for the acoustic contribution. Further, in the equipartition-regime valid for the temperatures and Fermi levels considered in the simulations, this acoustic contribution to the diagonal of the BTE matrix can be simplified to a inverse scattering time linear in temperature and energy of the electronic state (see Eq. C6 of Ref. \cite{Sohier2014}). For each Fermi level, we therefore perform a linear regression on the resistivity between $50$ K and $150$ K, to extract the acoustic phonon contribution (and the residual resistivity). Then, we assume the rest of the resistivity is due to optical phonons.

Regarding optical phonons, thanks to the isotropy of Dirac cones, it turns out that their contribution to the BTE is also diagonal if the electron-phonon matrix elements are isotropic (see the case of the sum of LO and TO phonons in Ref. \cite{Sohier2014}). Indeed, the off-diagonal terms are obtained via angular integrals over iso-energetic contours of the band structure, and in this case the integrand is proportional to the cosine of the scattered states angle, such that it integrates to zero.
In general, anisotropic electron-phonon matrix elements would simply lead to the coupling being weighted by the corresponding angular integral.
For the fitting procedure, we do not need to extract the exact values of the coupling's matrix elements. Instead, our fitting function must primarily capture the right temperature dependence of the resistivity associated to optical phonons with certain frequencies.

Thus in the electronic BTE we assume generic optical phonons with isotropic matrix elements, so that the diagonal elements of the BTE matrix are simply given by the following energy-dependent inverse scattering time:
\begin{equation}
    \frac{1}{\tau_{\nu}(\varepsilon)} = C_{\nu} C_0 \frac{1-f(\varepsilon \pm \hbar \omega_{\nu})}{1-f(\varepsilon)}  \left( n(\omega_{\nu}) + \frac{1}{2} \mp \frac{1}{2} \right) |\varepsilon \pm \hbar \omega_{\nu} |,
\end{equation}
where $\varepsilon$ is the energy of a given state (only energy matters because of isotropy), $\omega_{\nu}$ is the phonon frequency, and $f,n$ are Fermi-Dirac and Bose-Einstein occupations.
The $\pm$ signs correspond to absorption and emission, and the two terms are summed. $C_0 \approx 15.2$ eV$^{-1}$ ps$^{-1}$ ($1/100$ is atomic Rydberg units) is introduced such that the coupling constants $C_{\nu}$ are of the order of $1$.
% \textcolor{orange}{FM: I think we need to keep it in physical units. We do not need to divide by 100, but just express $C$ in the right units..So I would modify above equation}.
The acoustic and optical contributions to the BTE matrix are then added, giving the total inverse scattering time $1/\tau(\varepsilon)$ and the resistivity is obtained as
\begin{equation}
    \rho^{-1} = -\frac{e^2 v^2_{\rm F}}{2} \int d\varepsilon \rm{DOS}(\varepsilon) \tau(\varepsilon) \frac{\partial f}{\partial\varepsilon}(\varepsilon),
\end{equation}
where DOS is the density of states. Finally, we extract the optical coupling by fitting $\rho$ for both experiments and theory.

Practically, we assume that the optical contribution is due to two phonons, with fixed frequencies $\omega^{\rm BN}_{\rm ZO} = 800$ cm$^{-1}$ for BN's ZO phonon and $\omega^{\rm Gr}_{\rm A'_1} = 1225$ cm$^{-1}$ for graphene's $A'_1$ mode at the $\mathbf{K}$ point of the Brillouin zone. In principle, zone center LO/TO phonons from both graphene and BN may contribute too. However, given the $A'_1$ mode's strong coupling, the higher frequencies of the LO/TO phonons, and to avoid over-fitting, we do not include them here. Notice also that whatever contribution coming from these other high-energy phonons will likely produce some renormalization effect of the coupling to $\omega^{\rm Gr}_{\rm A'_1}$ phonon, rather than $\omega^{\rm BN}_{\rm ZO}$. Further, note that phonons with higher frequency will start to contribute to the resistivity at higher temperature. $\omega_{\rm A'_1} = 1225$ cm$^{-1}$ is already fairly large, and it barely starts to contribute below room temperature. Since our experiments stop at room temperature, we do not expect the experimental data to provide a reliable description of this phonon, much less the higher energy ones. While including a phonon at $\omega_{\rm K} = 1225$ cm$^{-1}$ is useful to adjust for an increase of resistivity close to room temperature, we shall refrain from drawing too many physical conclusions from the details of the results.

\begin{figure}
    \centering
    \includegraphics[width=0.98\linewidth]{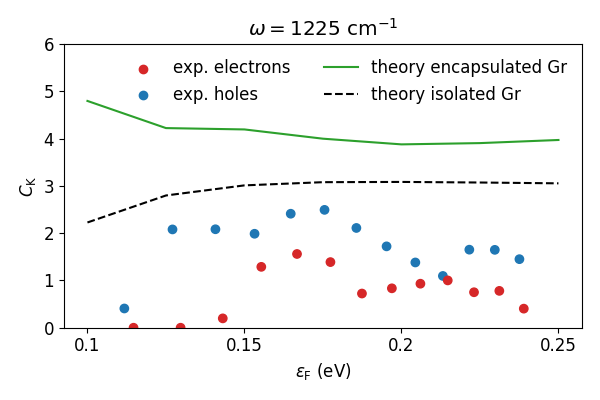}
    \caption{Coupling to the  graphene's $A_1$ mode at $\mathbf{K}$ fitted on the experimental resisitivities for electrons and holes, as well as simulated resisitivity in isolated graphene and encapsulated graphene.}
    \label{fig:C_Kfit}
\end{figure}

We call the fitting coupling strength parameters for BN's ZO phonons and graphene's $A_1$ mode at $\mathbf{K}$ as, respectively, $C_{\rm ZO}$ and $C_{\rm K}$. The former is plotted in the main text, while we report the latter here in Fig. \ref{fig:C_Kfit}.
First noting that they are all of the same order of magnitude, we will simply remark on their order. The calculations in encapsulated graphene give a larger coupling than for isolated graphene. This is perfectly reasonable, since encapsulated graphene contains the same optical phonons as the isolated graphene (graphene's $A_1$, LO, and TO phonons) plus BN's LO mode, that may renormalize $C_{\rm K}$ as anticipated.
Experiments consistently give slightly smaller couplings than both simulations. Given the roughness of the procedure, we are content with matching flat trends between theory and experiments.

\section{Additivity of resistivity}
\label{sec:additivity}

\begin{figure}
    \centering
    \includegraphics[width=0.98\linewidth]{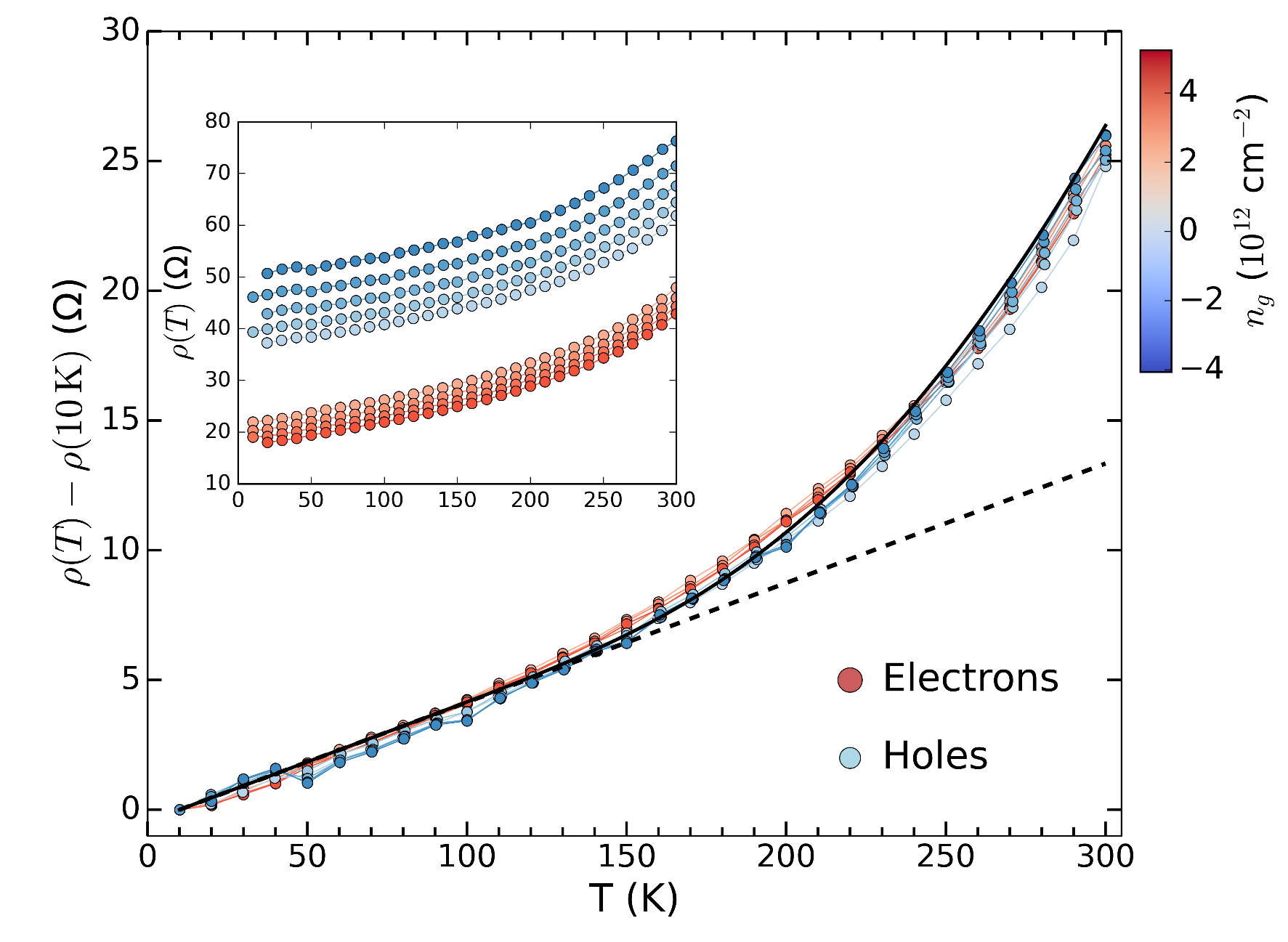}
    \caption{Same data as in the main text. They are fitted here using Eq. \ref{eq:rhowrong}. The dashed line is a guide to the eye, highlighting the linear regime with a slope of approximately 0.05 $\ohm$/K, and the black curve illustrates the trend of the full curve.}
    \label{fig:Boltzmann_fit}
\end{figure}

\begin{figure}
    \centering
    \includegraphics[width=0.98\linewidth]{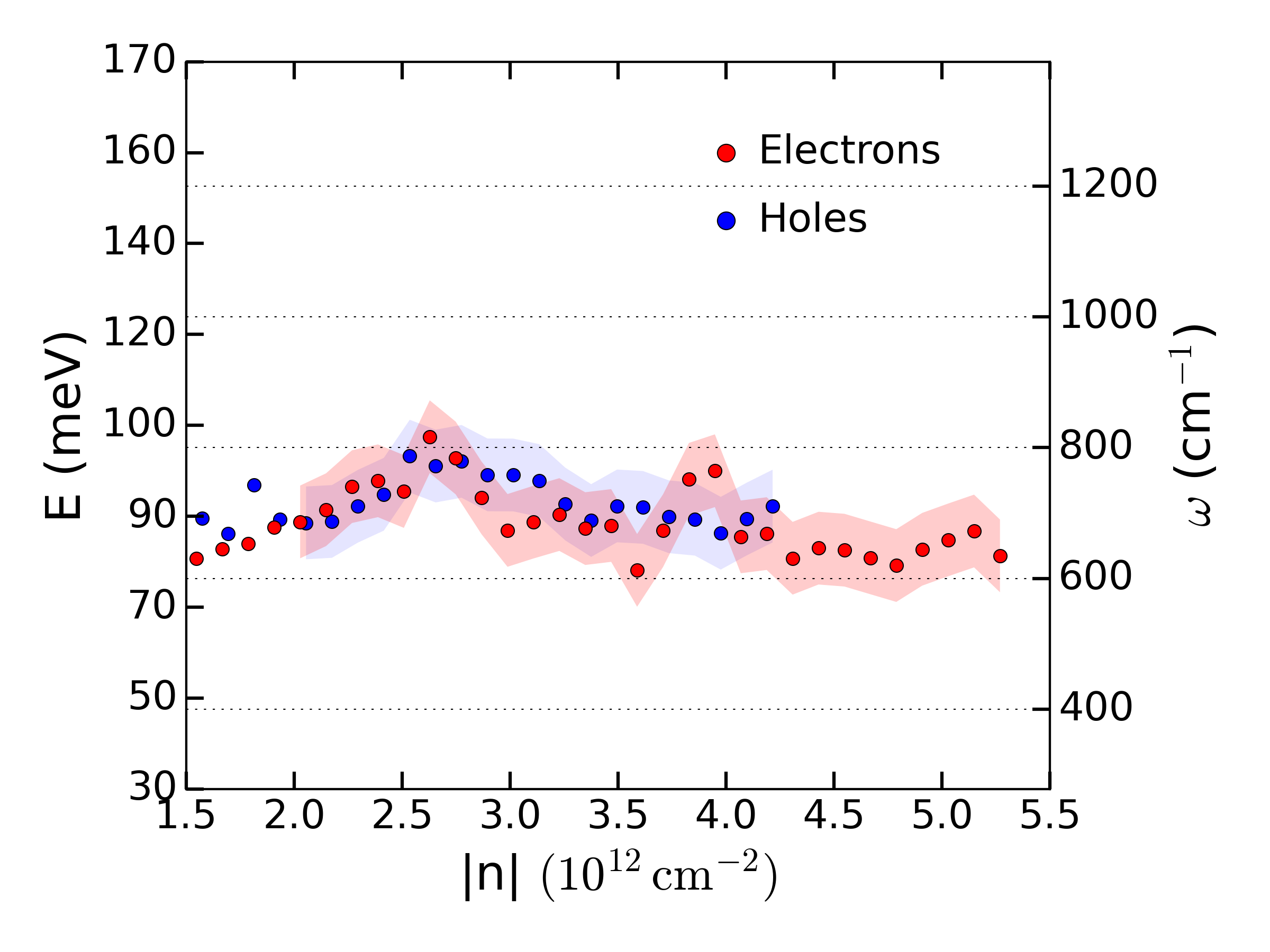}
    \caption{Phonon energy $E$ and frequency $\omega$ extracted by fitting the temperature-dependent resistivity using the phenomenological model Eq. \ref{eq:rhowrong}. Data are shown for electrons (red) and holes (blue). The extracted values of $E$ remain nearly constant and is systematically underestimated. }
    \label{fig:Boltzmann_fit_energy}
\end{figure}

Here we discuss a common practice in the literature \cite{Chen2008,You2019} which is to assume that the resistivities are additive and that the temperature dependence of the contribution from an optical phonon simply follows the associated Bose-Einstein occupation function. For graphene with acoustic phonons and one optical phonon, the resistivity would then be modeled as:
\begin{equation}
    \rho \sim \rho_0 + \alpha T + \frac{B}{e^{\hbar \Omega/k_BT}-1}.
    \label{eq:rhowrong}
\end{equation}
While this approximation is decent, and the resulting resistivity curves look somewhat similar (see Fig. \ref{fig:Boltzmann_fit}), we found that using this function for fitting is quite inaccurate.
In particular, if we leave the phonon frequency as a fitting parameter, we consistently get smaller values than we should (see Fig. \ref{fig:Boltzmann_fit_energy}).

We performed the following test. We solve the electronic BTE with acoustic phonons plus a single generic optical phonon as in \ref{app:fitProc}, with $\omega^{\rm BN}_{\rm ZO} = 800$ cm$^{-1}$ and $C_{\rm ZO} = 2 C_0$, and compute resistivity. We then fit the obtained resistivity with Eq. \eqref{eq:rhowrong}, with both $B$ and $\Omega$ as fitting parameters, finding $\Omega = 712$ cm$^{-1}$ and $B=1026$ Ohms. The value of $B$ is not directly comparable to $C_{\rm ZO}$. The significantly lower value of the phonon frequency, however, is a problem. Indeed, we rely on injecting known optical phonon frequencies in the fitting procedure. Clearly, this test indicates that to fit the ZO phonon properly with the additive resistivities, one should not fix $\Omega = \omega_{\rm ZO} = 800$ cm$^{-1}$, but rather some unknown renormalized effective value ($712$ cm$^{-1}$ in this particular case, but different in other conditions, e.g. different $\alpha$).
Thus, we consider the method of \ref{app:fitProc} to be much more reliable. Obviously, performing the above test with the method of \ref{app:fitProc}, it yields the correct values $\omega_{\rm ZO} = 800$ cm$^{-1}$ and $C_{\rm ZO} = 2 C_0$.

\bibliography{VEDcBTE_BNGr_ZO}